\documentclass[preprint,aip]{revtex4-1}
\usepackage{graphicx,xcolor}
\usepackage{dcolumn}
\usepackage{bm}
\usepackage{amsfonts,amssymb,amsmath,geometry, url,siunitx}
\usepackage{gensymb}
\usepackage{graphicx}
\usepackage{hyperref}
\usepackage{lineno}
\usepackage{hyperref}
\hypersetup{
   colorlinks=true,    
   linkcolor=blue,       
   citecolor=blue,       
   filecolor=black,        
   urlcolor=blue,        
   linktoc=page            
}

\begin{document}

\title{GaN Nanowall Network: A new possible route to obtain efficient p-GaN and enhanced light extraction } %

\author{Sanjay Kumar Nayak}
\email{sanjaynayak@jncasr.ac.in} 
\affiliation{Chemistry and Physics of Materials Unit \\ Jawaharlal Nehru Centre for Advanced Scientific Research (JNCASR), Bangalore-560064, India}
\author{Mukul Gupta}%
 \email{mgupta@csr.res.in}
\affiliation{UGC-DAE Consortium for Scientific Research, Khandwa Road, Indore-452017, India}
\author{S.M. Shivaprasad}
\email{smsprasad@jncasr.ac.in}
\affiliation{Chemistry and Physics of Materials Unit \\ Jawaharlal Nehru Centre for Advanced Scientific Research (JNCASR), Bangalore-560064, India}
\affiliation{International Centre for Materials Science \\ Jawaharlal Nehru Centre for Advanced Scientific Research (JNCASR), Bangalore-560064, India}
\date{\today}

\begin{abstract}
 We demonstrate that GaN formed in a Nanowall Network (NwN)  morphology can  overcome fundamental limitations in optoelectronic devices, and enable high light extraction  and  effective Mg incorporation for  efficient  p-GaN. We report the growth of Mg doped GaN Nanowall network (NwN) by plasma assisted molecular beam epitaxy (PA-MBE) that is  characterized by Photoluminescence (PL) spectroscopy, Raman spectroscopy, high-resolution X-ray diffraction (HR-XRD), X-ray photoelectron spectroscopy (XPS) and Secondary ion mass spectroscopy (SIMS). We record a photo-luminescence enhancement ($ \approx $3.2 times) in lightly doped GaN as compared to that of undoped NwN. Two distinct (and broad) blue luminescence peaks appears at 2.95 and 2.7 eV for the heavily doped GaN (Mg $>10^{20}$ atoms $cm^{-3}$),  of which the 2.95 eV peak is sensitive to annealing  is observed. XPS and SIMS measurements estimate the incorporated Mg concentration to be $10^{20}$ atoms $cm^{-3}$ in GaN NwN morphology, while retaining its band edge emission at $\approx$ 3.4 eV. A higher Mg accumulation towards the GaN/Al$_2$O$_3$ interface as compared to the surface was observed from SIMS measurements.
\end{abstract}

\maketitle

\section{Introduction}
Gallium nitride (GaN) is one of the most preferred semiconductor material for light-emitting diodes (LEDs)\cite{Dupuis2008}, high electron mobility transistor (HEMT)\cite{Sheppard1999}, lasers\cite{Lu2008}, dilute magnetic semiconductors (DMS)\cite{Sasaki2002}, solar cells\cite{Sheu2009,Kuwahara2010}, photo electrochemical water splitting\cite{Wang2011a}, biosensors\cite{Pearton2004a}, and  space applications\cite{Muraro2010,Miwa2011}. Such widespread applications of GaN are due to its direct band gap, bandgap tunability\cite{Kuykendall2007}, high mobility, chemical and thermal stability  and better internal quantum efficiency\cite{Pimputkar2009} for light emission as compared to other semiconducting materials such as GaAs and ZnO. Despite great progress in the synthesis of high-quality thin-films with low dislocation density, the overall efficiency of GaN-based LED remains low  due to lower light extraction efficiency (LEE), caused by a large mismatch in refractive indices between GaN and the ambient\cite{Wiesmann2009a}. 

Several methods has been proposed \cite{Choi2003,David2007,Bilousov2014} to enhanced light extraction efficiency from these materials, among which  the use of  nano-porous\cite{Soh2013,Yang2008} structure has shown great promise. Generally, the synthesis of porous structures are achieved by chemical or ion bombardment etching, which unfortunately are prone to introduce defects and impurity states in the materials and thus leading to degradation of crystal integrity, which consequently lowers  device performance. Thus, fabrication of the porous structure by the bottom-up approach, by controlling the kinetics of growth enables avoidance of such pitfalls. Previously, we have shown that by controlling growth parameters such as substrate temperature and III-V ratio, the GaN NwN can be formed directly on c-plane  sapphire\cite{Kesaria2011,Thakur2015}. We have reported that the NwN exhibits superior structural and optical properties as compared to that  of flat GaN thin films. Our experimental and  simulation based on finite difference time domain (FDTD),  findings showed that the observed  high PL intensity from GaN  NwN  is not only due to dislocation filtering but also due to the geometry of its unique surface morphology, which facilitates  escape of the generated photons to the ambient by reducing total internal reflection\cite{Nayak2016a}.\\\\
It is well known that n-type GaN can be obtained much easily as compared to p-type. Magnesium (Mg) is the most successfully used dopant to make p-type GaN, where it substitutes Ga atom in the GaN lattice resulting in an acceptor state in its electronic structure. However, the higher ionization energy ($\approx$200 meV) of Mg \cite{Zhang2010}, and a high unintentional  n-doping of intrinsic GaN, makes it difficult to obtain highly p-doped GaN. Thus, a very high concentration of Mg ($>10^{20} cm^{-3}$)  is required \cite{Cimpoiasu2006} for obtaining significant and useful p- doping. Typical Mg dopant concentrations of $10^{17} -10^{19} cm^{-3}$ have been incorporated in planar GaN films\cite{Ptak2001,Myers2001} while higher Mg incorporation is seen to form  defects such as N-vacancy and Mg- interstitials, and $Mg_{Ga}V_{N}$ like clusters. In addition, there can be polarity inversion in the film that can also lead to the degradation of its crystal structure, which results in  poor optical properties. Moreover, these defects and complexes may result in self-compensation in p-GaN\cite{Smorchkova2000,Miceli2016}. Experiments based on X-ray photoelectron spectroscopy (XPS) and Secondary ion mass spectroscopy (SIMS) measurements of GaN flat film show that concentration of Mg on the surface is higher than in the bulk\cite{Nakano2002,Hashizume2003,Cheng1999,Romano2001}suggesting that, GaN with higher surface area may enable higher incorporation of Mg. Since NwN with its porous structure has a very high surface to volume ratio,  we considered it to be a potential candidate for achieving higher incorporation of Mg. In this report we study the role of Mg concentration on morphology, crystal structure and optical properties of the GaN NwN.

\section{Experimental Details}
The GaN NwN films are grown on bare c-plane of sapphire ($\alpha$-Al$_2$O$_3$) under nitrogen rich conditions by using radio frequency plasma assisted molecular beam epitaxy system (RF-PAMBE, SVTA-USA), operating at a base pressure of $3\times 10^{-11}$ torr. The detailed procedure of substrate preparation can be found elsewhere\cite{Kesaria2011f}. The temperature of Gallium (Ga) effusion cell is kept at 1030 $^o$C. A constant nitrogen flow rate of 8 sccm (standard cubic centimeter per minute), substrate temperature of 630 $^o$C, plasma forward power of 375 W and growth duration of 4 hours were maintained for all the films. Mg flux was varied by controlling Mg k-cell temperature from 300 $^o$C to 360 $^o$C in steps of 20 $^o$C. The flux of Mg and Ga were obtained from the beam equivalent pressure (BEP) and are tabulated in Table 1. Surface structural evolution was monitored \textit{in-situ} by reflection high energy electron diffraction (RHEED) and the morphology was determined \textit{ex-situ} by a field emission scanning electron microscope (FESEM). Structural quality of the films is determined by a high-resolution X-ray diffractometer (HR-XRD, Discover D8 Bruker) with a Cu K$_\alpha$ X-ray source of wavelength of 1.5406 \AA. Optical properties of the films were studied by photo-luminescence spectroscopy (PL, Horiba Jobin Yvon) using a Xenon lamp source with 325 nm excitation, Raman spectroscopy with Ar laser of wavelength 514 nm is performed in the back scattering geometry.  Quantification of Mg incorporated in the film is done by \textit{ex-situ} X-ray Photoelectron Spectroscopy (XPS) using Omicron EA 125 spectrometer and Al−K$_\alpha$ (1486.7 eV) source, with a relative composition detection better than 0.1\%. Before performing XPS measurements, GaN NwN films were sputter cleaned by optimized low energy (0.5 keV, 2 $\mu$A) Ar$^+$ ions, to remove physisorbed adventitious carbon and oxygen resulting from atmospheric exposure, without affecting the crystalline quality and composition of the film.  The depth dependence of Mg distribution has been studied by secondary ion mass spectroscopy (SIMS) using O$^{+2}$ ion beam of 3 keV.

\begin{table*}[t]
 \caption{\label{tab}  Mg flux rate } 
\begin{ruledtabular}
   \centering
 \begin{tabular}{ ccccccccc rrrrrrrrrrr }
 {Sample Name}  & {Mg-k cell temp ($^o$ C)}   & {BEP (torr)}  & {Flux ($cm^{-2}s{^-1}$)}  & {Mg:Ga}   \\
  
 \hline \
{A} & {-} &{-} & {-} & {0}  \\
{B} & {300}&{$2.9\times10^{-10}$ } & {$8.6\times10^{10}$ }&{0.0017}   \\
{C} &{320 } &{$7.1\times10^{-9}$} & {$2.0\times10^{12}$}& {0.0393}  \\
{D} & {340 } &{$1.1\times10^{-8}$} & {$3.1\times10^{12}$ }& {0.0622} \\
{E} & {360 } &{$2.0\times10^{-8}$} & {$5.6\times10^{12}$ }& {0.1102} \\
 
\end{tabular}
\end{ruledtabular}
\end{table*}

\section{Results and Discussion}
\begin{figure}[h]
   \centering
       \includegraphics[width=15cm]{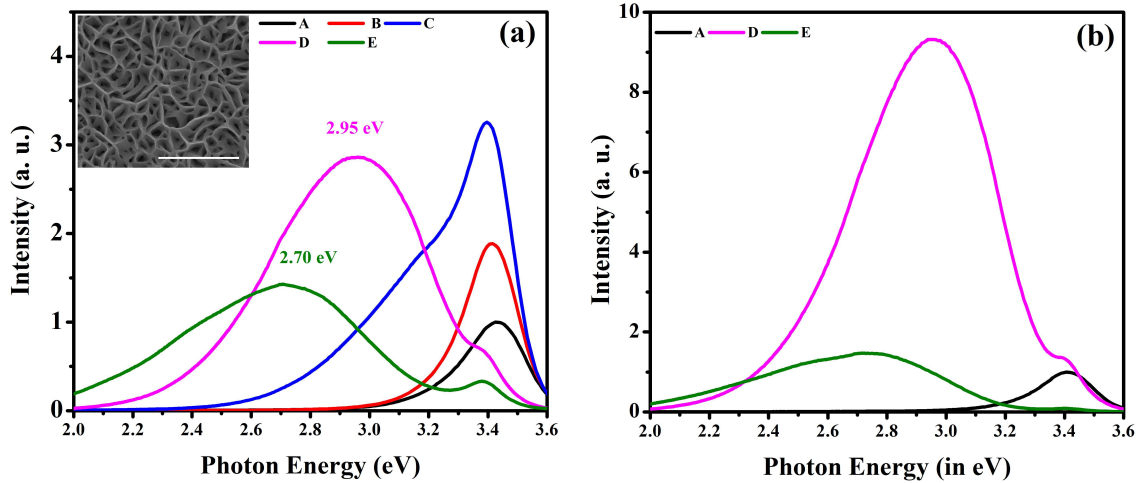}
    \caption{(a)shows photoluminescence (PL) spectra of the undoped and Mg doped GaN NwN. Inset of the figure represents SEM image of sample E with the scale bar of 500 nm. (b) shows the PL spectra of sample D and E after annealing at 800 $^o$C in presence of N$_2$ for 20 minutes.}
    \label{PL}
 
\end{figure}

In our previous study we have shown that the unique topography of GaN NwN structure results in  superior optical properties \cite{Nayak2016a}. In the present study we control Mg incorporation in NwN  while the $Mg:Ga$ ratio was maintained in the range of $0.0017 < Mg:Ga < 0.1102$. The surface morphology of the grown films are obtained by SEM imaging and found to have   similar morphology  independent of the doping concentration. The typical SEM image of sample E (grown with maximum Mg flux rate) is shown as inset in Fig.\ref{PL}.  

PL and Raman spectroscopy techniques have been intensively employed in the past to study the effect of Mg doping on the optical properties of GaN\cite{Reshchikov2014,Kirste2013,Gelhausen2004,Harima2002}.  In the literature, commonly observed luminescence peaks of Mg doped GaN appears at $\approx$ 3.27 eV i.e. Ultra-violet luminescence (UVL) and a broad  $\approx$ 2.7-2.9 eV blue luminescence (BL)\cite{Reshchikov2014}. Fig.\ref{PL} shows PL spectra of samples with different Mg doping concentrations (see Table \ref{tab}). The PL intensities of samples B, C, D and E are shown relative to that of undoped GaN NwN (sample A). We observe significant changes in the luminescence spectra of doped films in comparison to that of undoped film. The near band edge (NBE) is dominant in undoped (A) and lightly doped (B and C) samples and its intensity increases with the increasing Mg:Ga flux. However, in the heavily doped GaN NwN samples (D and E),  BL is the dominant emission. We record  2 times and 3.25 times increase of  PL intensity  for  sample B and  sample C, respectively, w.r.t. sample A. An increase in PL intensity at lower Mg incorporation has been previously attributed to screening of polarization induced field by Mg\cite{Zhang2013a} which enable efficient overlapping of \textit{e-h} wave-function, resulting the higher recombination of them. However, the samples D and E grown at higher Mg flux have their NBE intensity significantly quenched while the intensity of broad BL peaks increases. We note that in sample C, unlike in un-doped A and sample B, a shoulder peak is observed at $\approx$ 3.2 eV. The peak is identified as donor-acceptor pair (DAP) luminescence originating either from the transition from shallow donor to a shallow acceptor level or due to Mg-H complex, while other peaks in the range of 3.1-3.2 eV are due to conduction band to shallow acceptor level transitions\cite{Reshchikov1999}. In the present case, since the films were grown using MBE, the formation of Mg-H complex is less probable; therefore, we attribute this shoulder peak to DAP transition.   

In cases of higher Mg flux (D and E); we observe reduced intensity in NBE emission from the Mg doped GaN NwN. In the case of sample D, the peak at 2.95 eV is dominant and a reduced NBE appears at 3.4 eV. For sample E, the NBE emission has further diminished and a broad peak at 2.7 eV becomes dominant. The origin of such BL peaks is highly debated\cite{Monemar2014} while some recent experimental \cite{Gelhausen2004,kaufmann1998} and Density Functional Theory (DFT)\cite{Qimin2012,Buckeridge2015} studies suggest some understanding. Usually the 2.9 eV peak appears in heavily Mg doped GaN and is regarded as the signature of Mg doped GaN\cite{Gelhausen2004}. While Kaufmann \textit{et al.}\cite{Kaufmann1999} attribute this peak to  vacancy complex such as $Mg_{Ga}V_N$,  Akasaki \textit{et al.}\cite{Koide2002} to the hydrogen-related deep donor to Mg acceptor states.   Some reports suggest  the BL (2.7-2.9 eV) is due to the transition from CBM to the deep acceptor, which appears due to hole localization\cite{Lyons2012}. It has been proposed theoretically that\cite{Qimin2012,Lyons2012}, the relaxation pattern  (local strain behaviour) of neutral and charged $Mg_{Ga}$ configuration is different from each other, which results  in the creation of a deep acceptor state 0.54 eV above the $Mg_{Ga}$ related shallow acceptor state, which lies 260 meV above the bulk VBM. The appearance of this peak confirms the incorporation of Mg in GaN. It can be clearly seen from the PL spectra (see Fig.\ref{PL}) that the BL peak is intense and broad. The cause of broadening in such luminescence peaks can be understood from the configuration co-ordinate diagram shown in Ref.\onlinecite{Reshchikov2014, Lyons2012}. Van de Walle \textit{et al.} \cite{Lyons2012} estimated that the energy difference between the BL and the zero phonon line is around 0.54 eV, which results in the broadening of BL in PL spectra.

Thus, the peaks observed at 3.43, 3.42, 3.41, 3.40 and 3.39 eV are attributed as NBE for undoped GaN and Mg doped GaN by using Mg flux of $8.6\times10^{10}$, $2.0\times10^{12}$, $3.1\times10^{12}$ and $5.6\times10^{12}$ $cm^{-2}s^{-1}$, respectively. The  results suggest that Mg doping in GaN yields a red-shift in NBE which is consistent with earlier reports\cite{Sui2011}. To study the impact of annealing on BL of Mg doped GaN, we annealed sample D and E at 800 $^o$C for 20 minutes in the presence of $N_2$. The PL spectra (see Fig.\ref{PL}(b)) of annealed samples showed a large increase in the intensity of 2.95 eV peak (for sample D) whereas the intensity of 2.7 eV peak (for sample E) did not change significantly, which indicates that origin of both luminescence peaks is different. The origin of BL in Mg doped GaN also shows that the Mg concentration\cite{Reshchikov2005d} in GaN is of the order of $10^{19}-10^{20}$ atoms $cm^{-3}$  .
\begin{figure}[h]
   \centering
       \includegraphics[width=8cm]{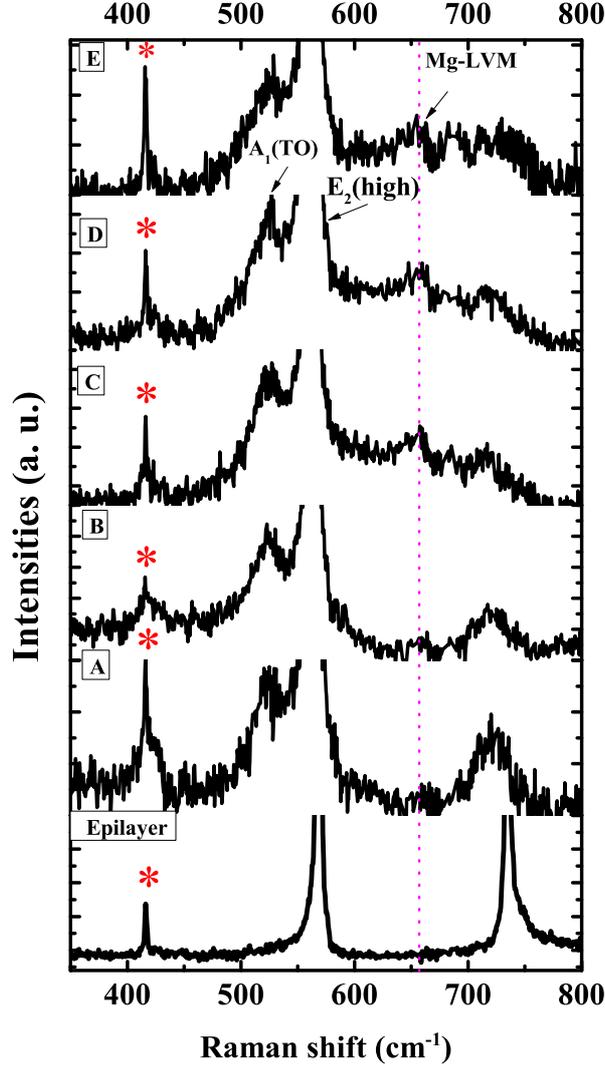}
    \caption{shows Raman spectra of the samples studied here. `` * " represents phonon mode from sapphire. The appearance of Mg related LVM can be seen with the increase in Mg:Ga flux. The dotted line is drawn at 657 $cm^{-1}$ to ease the visualisation.}
    \label{raman}
\end{figure}
\par
\begin{figure}[h]
   \centering
       \includegraphics[width=8cm]{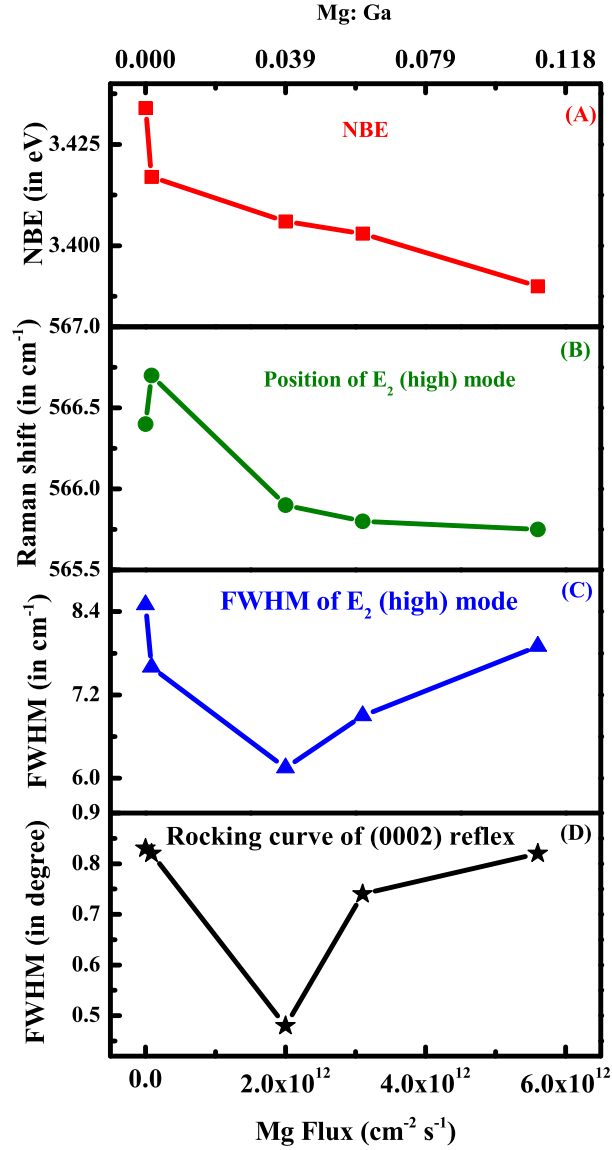}
    \caption{(A), (B), (C), (D) represent  NBE, FWHM of Raman E$_2$ (high) mode, position of Raman E$_2$ (high) mode  and XRD rocking curve of (0002) reflex as a function of Mg flux (or Mg: Ga).}
    \label{result}
\end{figure}

To study the impact of dopants on the local strain in the grown films, the local vibrational modes (LVM) are monitored by Raman spectroscopy and are shown in  Fig.\ref{raman}. The LVM of Mg doped GaN\cite{Harima2002,Kirste2013} appears at $\approx$ 657 $cm^{-1}$. As Mg replaces Ga, a compressive strain is introduced which shift the E$_2$ (high) mode towards higher energies. The E$_2$ (high) mode being a non-polar mode is suitable for the study of in-plane strain\cite{Harima2002}.  To observe morphology induced changes, we have also plotted the Raman spectra from a flat 3$\mu$m thick GaN epilayer on c-sapphire. Along with the allowed modes the two geometrically forbidden modes such as E$_1$ (TO) and A$_1$(TO) are  observed in all the GaN NwN samples, due  to the scattering off the sidewalls of the porous structure\cite{Thakur2015,Williamson2004a}. We clearly observe (see Fig.\ref{raman}) the presence of LVM at 657 $cm^{-1}$, which is the Mg-N stretching bond. The LVM peak is absent for both GaN epilayer and undoped NwN, but increases with higher Mg flux. The variation of position and FWHM of E$_2$ (high) mode with Mg flux as well as Mg to Ga ratio for all the studied samples are shown in Fig.\ref{result}. As it can be seen, the position of E$_2$ (high) mode for sample B shifts towards the higher energy, as expected but for other samples  this mode shifts towards lower energy. As, Mg in GaN usually induces compressive strain because of its higher ionic radius, the  E$_2$ (high) mode shifts towards the high-frequency. However, if  defect complexes or different charged states, other than $Mg_{Ga}^0$  , such as $Mg_{Ga}^-$ are formed, than the strain behaviour varies. This could be the possible reason for a shift of phonon frequency towards the lower energy, since, now a tensile strain is developed in the structure due to Mg incorporation\cite{Kirste2013}. The FWHM of E$_2$ (high) mode shows a large change from the highest value of 8.5 $cm^{-1}$ in GaN NwN to 7.6, 5.6, 7.2, and 7.9 $cm^{-1}$ for samples B, C, D and E, respectively, due to the incorporation of different defects\cite{Kirste2013} in the films.

Strain is introduced in the system by not only Mg replacing Ga in the lattice, but also due to the formation of point and extended defects and impurity complexes, which can also modify the lattice parameter\cite{Moram2009,Kirste2013}. To study the impact of Mg incorporation on the crystal structure of the NwN, HR-XRD study has been carried out. We record both 2$\theta-\Omega$ and $\omega$-scans, which provide information about stress and crystal quality of the samples, respectively. We compare the acquired results with those of an undoped GaN NwN. We have reported earlier that\cite{Thakur2015}, undoped GaN NwN possesses very low strain as compared to a flat epitaxial thin film. For Mg-doped GaN NwN we observe a complex strain behaviour. The estimated ‘c’ lattice parameter from HR-XRD for samples A, B, C, D and E are 5.1915, 5.1926, 5.1913, 5.1909 and 5.1956 \AA, respectively which varies from that of the un-doped sample A by 0.02\%, -0.004\%, -0.011\%, 0.08\%, respectively. The lower values of the lattice parameters (for sample C and D) and the shifts of E$_2$ (high) mode suggest the presence of charged defect or defect complexes formed due to Mg doping. However, for sample E, the E$_2$(high) mode shifts towards lower frequency, but the ‘c’ lattice parameter increases which suggests that interstitial type of defects are incorporated. The FWHM of the rocking curves of (0002) reflex for the samples  shown in Fig.\ref{result} (D) displays an ``U" type behaviour, with increasing Mg:Ga flux. The broadening of XRD in NwN is mainly due to the mosaicity that arises due to the misalignment of the nanowalls as seen earlier\cite{Nayak2016a,Poppitz2014b}. We observed a reduced FWHM value for sample C, but it increases  for high Mg: Ga flux (sample D and E),  due to the formation of defects.

 \begin{figure}[h]
   \centering
       \includegraphics[width=8cm]{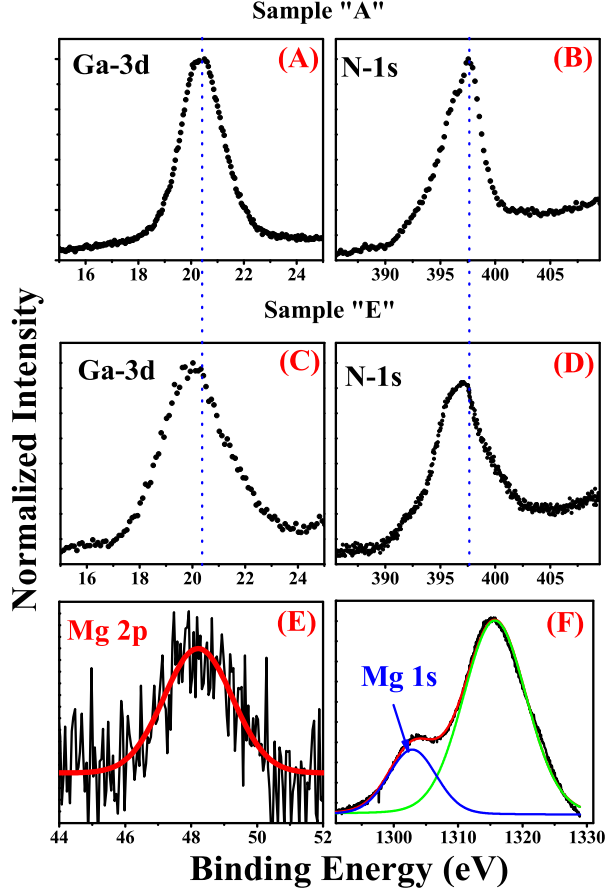}
    \caption{(A) and (B) shows Ga-3d and N-1s core level spectra of undoped GaN NwN (sample A). Fig.\ref{xps} (C), (D), (E) and (F) represent Ga-3d, N-1s, Mg-2p and Mg-1s core level spectra of Mg doped GaN NwN (sample E).}
    \label{xps}
\end{figure}

We performed XPS measurement of samples A and E to study the electronic structure variation  and to quantify Mg in sample E. Fig.\ref{xps} (A) and (B) represents Ga-3d and N-1s core level spectra of sample A, whereas (C), (D), (E) and (F) represent Ga-3d, N-1s, Mg-2p, Mg-1s core level spectra of sample E, respectively.  As can be clearly seen from the figure the width of Ga-3d and N-1s core levels of doped sample is relatively higher than that of un-doped NwN. The peak position of Ga-3d and N-1s core levels of E are shifted by 0.3 eV and 0.5 eV, respectively, as compared to sample A suggesting a downward shift of Fermi level of p-GaN and unintentionally n-doped GaN, due to  Mg incorporation in the GaN NwN. \cite{yongdeuk2015}. We estimate the quantity of Mg incorporation in sample E by the relation \\
\begin{center}
                            $C_{Mg} (\%) = \dfrac{I_{Mg}/ASF_{Mg}}{\sum{I}_i /ASF_{i}} \times100 $
\end{center}
where $C_{Mg}$, $I_{Mg}$, $ASF_{Mg}$ represents is the fraction of Mg,  intensity of Mg-1s peak and its atomic sensitivity factor, while $I_i$ and $ASF_i$ are intensity and atomic sensitivity factors of the other constituent elements. We used atomic sensitivity factors for Ga-2p$_{3/2}$, N-1s and Mg-1s as 3.720, 0.477 and 3.500, respectively. The estimated value of Mg in the sample is ≈ 3.3\% which corresponds to a Mg concentration of $\approx 9\times10^{20}$ $cm^{-3}$, which is quite consistent with our predictions from PL measurements .
\par
\begin{figure}[h]
   \centering
       \includegraphics[width=8cm]{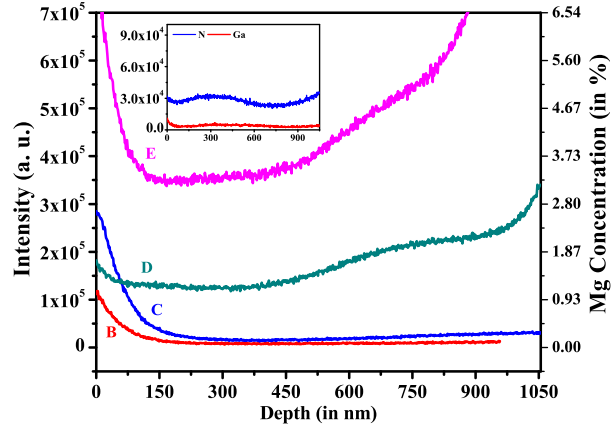}
    \caption{shows the depth profile of Mg as a function of depth. Inset of the figure shows depth profile of  Ga and N of the sample D.}
    \label{sims}
\end{figure}

Further, to understand the Mg distribution profile and estimate the Mg concentration in the grown films, we carried out SIMS measurements on all samples as shown in Fig.\ref{sims}. We estimate the Mg concentrations of remaining  samples  by using the relation, 
\begin{center}
                                                        $ I_{A}= Y.f.I_P.C_A.P^{\pm} $ 
\end{center}

where $I_{A}$ and $C_A$ represent secondary ion intensity and its concentration. $I_P$, Y and $P_{\pm}$ represent primary ion current, sputtered yield and ionization probability of the secondary ion. Within the very narrow regime, the impurity concentration is linearly proportional to the secondary ion intensity. The Mg concentration of all films are shown in Fig.\ref{sims}.  As discussed earlier, Mg is observed to segregate to the surface in Mg doped  flat GaN thin films\cite{Nakano2002,Hashizume2003,Cheng1999,Romano2001}. Our results (see Fig.\ref{sims}) show that the Mg segregates both  at surface of the NwN and GaN/$Al_2O_3$ interface for samples D and E.  Arbitrary intensity proportional to the counts measured is shown along Y-axis, but the alternate Y-axis is shows in terms of \% composition obtained by XPS measurement.  For heavily doped sample E, Mg accumulates  more at the interface. However, depth profile of Mg in samples B and C   shows Mg segregation only at the surface and depletes monotonically.   To further confirm the accumulation of Mg at the interface, we obtained a depth profile of two different isotopes of Mg such as $^{24}Mg$ and $^{25}Mg$ (not shown here) and we observe similar behavior for both the isotopes. To further  verify that this profile is not due to the unique geometry of the NwN, we recorded N and Ga depth profiles of all the samples. As a typical case, the N and Ga depth profile of  sample D  is presented as inset of Fig.\ref{sims}. The profile shows  a fairly constant Ga and N distribution along the depth. \\

It has been proposed theoretically that, Mg incorporation is easier in Ga-polar surface than in N-polar surface\cite{Sun2006a} and also such doping can change the polarity of the film from Ga-polar to N-polar\cite{Romano2000,Sun2006a}. Convergence Beam Electron Diffraction (CBED) measurements (not shown here)  on the  un-doped GaN NwN have shown that these films are  Ga-polar in nature. Thus, at the initial stage of that growth Mg incorporation is higher, while with growth time the change in polarity of the crystal may be the cause for the  lower incorporation of Mg.  We observe that the Mg concentration in the films towards the interface is 1.6 and 1.9 times higher as compared to that at the surface for the samples D and E, respectively.  

\section{Conclusions}
In conclusion, we have studied the structural, optical and dopant  distribution in  Mg-doped GaN NwNs. We find 3.2 times enhancement in NBE  for sample with low Mg flux (Mg:Ga $<$ 0.04), while higher Mg flux (Mg:Ga $>$ 0.06)  quenches NBE and increases   blue luminescence. Thermal annealing in presence of N$_2$ increases intensity of BL (2.95 eV) and does not change the 2.7 eV peak. We observe, with increasing Mg:Ga flux, the E$_2$ (high) mode shift towards the lower frequency indicating a change in local strain from compressive to tensile.  XPS along with SIMS reveals that more than $10^{20}$ $cm^{-3}$ Mg atoms can be incorporated in the sample due to its unique morphology with very high  surface to volume ratio. Higher light extraction capability along-with higher incorporation of Mg, in the NwN morphology suggests that these GaN films  can be used  to fabricate  p-type GaN for GaN/InGaN based quantum well structures for LEDs and LDs, with enhanced efficiency.

\section*{Acknowledgment}
The authors  thank Professor C. N. R. Rao for his support and guidance.  SKN acknowledges DST for  a Senior Research Fellowship and JNCASR for facilities.
\clearpage
\bibliographystyle{aipnum4-1}
\bibliography{Mg} 

\begin{thebibliography}{53}%
\makeatletter
\providecommand \@ifxundefined [1]{%
 \@ifx{#1\undefined}
}%
\providecommand \@ifnum [1]{%
 \ifnum #1\expandafter \@firstoftwo
 \else \expandafter \@secondoftwo
 \fi
}%
\providecommand \@ifx [1]{%
 \ifx #1\expandafter \@firstoftwo
 \else \expandafter \@secondoftwo
 \fi
}%
\providecommand \natexlab [1]{#1}%
\providecommand \enquote  [1]{``#1''}%
\providecommand \bibnamefont  [1]{#1}%
\providecommand \bibfnamefont [1]{#1}%
\providecommand \citenamefont [1]{#1}%
\providecommand \href@noop [0]{\@secondoftwo}%
\providecommand \href [0]{\begingroup \@sanitize@url \@href}%
\providecommand \@href[1]{\@@startlink{#1}\@@href}%
\providecommand \@@href[1]{\endgroup#1\@@endlink}%
\providecommand \@sanitize@url [0]{\catcode `\\12\catcode `\$12\catcode
  `\&12\catcode `\#12\catcode `\^12\catcode `\_12\catcode `\%12\relax}%
\providecommand \@@startlink[1]{}%
\providecommand \@@endlink[0]{}%
\providecommand \url  [0]{\begingroup\@sanitize@url \@url }%
\providecommand \@url [1]{\endgroup\@href {#1}{\urlprefix }}%
\providecommand \urlprefix  [0]{URL }%
\providecommand \Eprint [0]{\href }%
\providecommand \doibase [0]{http://dx.doi.org/}%
\providecommand \selectlanguage [0]{\@gobble}%
\providecommand \bibinfo  [0]{\@secondoftwo}%
\providecommand \bibfield  [0]{\@secondoftwo}%
\providecommand \translation [1]{[#1]}%
\providecommand \BibitemOpen [0]{}%
\providecommand \bibitemStop [0]{}%
\providecommand \bibitemNoStop [0]{.\EOS\space}%
\providecommand \EOS [0]{\spacefactor3000\relax}%
\providecommand \BibitemShut  [1]{\csname bibitem#1\endcsname}%
\let\auto@bib@innerbib\@empty
\bibitem [{\citenamefont {Dupuis}, \citenamefont {Krames},\ and\ \citenamefont
  {Member}(2008)}]{Dupuis2008}%
  \BibitemOpen
  \bibfield  {author} {\bibinfo {author} {\bibfnamefont {R.~D.}\ \bibnamefont
  {Dupuis}}, \bibinfo {author} {\bibfnamefont {M.~R.}\ \bibnamefont {Krames}},
  \ and\ \bibinfo {author} {\bibfnamefont {S.}~\bibnamefont {Member}},\
  }\href@noop {} {\bibfield  {journal} {\bibinfo  {journal} {J. Light.
  Technol.}\ }\textbf {\bibinfo {volume} {26}},\ \bibinfo {pages} {1154}
  (\bibinfo {year} {2008})}\BibitemShut {NoStop}%
\bibitem [{\citenamefont {Sheppard}\ \emph {et~al.}(1999)\citenamefont
  {Sheppard}, \citenamefont {Doverspike}, \citenamefont {Pribble},
  \citenamefont {Allen}, \citenamefont {Palmour}, \citenamefont {Kehias},\ and\
  \citenamefont {Jenkins}}]{Sheppard1999}%
  \BibitemOpen
  \bibfield  {author} {\bibinfo {author} {\bibfnamefont {S.~T.}\ \bibnamefont
  {Sheppard}}, \bibinfo {author} {\bibfnamefont {K.}~\bibnamefont
  {Doverspike}}, \bibinfo {author} {\bibfnamefont {W.~L.}\ \bibnamefont
  {Pribble}}, \bibinfo {author} {\bibfnamefont {S.~T.}\ \bibnamefont {Allen}},
  \bibinfo {author} {\bibfnamefont {J.~W.}\ \bibnamefont {Palmour}}, \bibinfo
  {author} {\bibfnamefont {L.~T.}\ \bibnamefont {Kehias}}, \ and\ \bibinfo
  {author} {\bibfnamefont {T.~J.}\ \bibnamefont {Jenkins}},\ }\href {\doibase
  10.1109/55.753753} {\bibfield  {journal} {\bibinfo  {journal} {IEEE Electron
  Device Lett.}\ }\textbf {\bibinfo {volume} {20}},\ \bibinfo {pages} {161}
  (\bibinfo {year} {1999})}\BibitemShut {NoStop}%
\bibitem [{\citenamefont {Lu}\ \emph {et~al.}(2008)\citenamefont {Lu},
  \citenamefont {Kao}, \citenamefont {Chen}, \citenamefont {Kao}, \citenamefont
  {Kuo},\ and\ \citenamefont {Wang}}]{Lu2008}%
  \BibitemOpen
  \bibfield  {author} {\bibinfo {author} {\bibfnamefont {T.~C.}\ \bibnamefont
  {Lu}}, \bibinfo {author} {\bibfnamefont {T.~T.}\ \bibnamefont {Kao}},
  \bibinfo {author} {\bibfnamefont {S.~W.}\ \bibnamefont {Chen}}, \bibinfo
  {author} {\bibfnamefont {C.~C.}\ \bibnamefont {Kao}}, \bibinfo {author}
  {\bibfnamefont {H.~C.}\ \bibnamefont {Kuo}}, \ and\ \bibinfo {author}
  {\bibfnamefont {S.~C.}\ \bibnamefont {Wang}},\ }\href {\doibase
  10.1109/CLEO.2008.4551202} {\bibfield  {journal} {\bibinfo  {journal} {Appl.
  Phys. Lett.}\ }\textbf {\bibinfo {volume} {92}},\ \bibinfo {pages} {141102}
  (\bibinfo {year} {2008})}\BibitemShut {NoStop}%
\bibitem [{\citenamefont {Sasaki}\ \emph {et~al.}(2002)\citenamefont {Sasaki},
  \citenamefont {Sonoda}, \citenamefont {Yamamoto}, \citenamefont {Suga},
  \citenamefont {Shimizu}, \citenamefont {Kindo},\ and\ \citenamefont
  {Hori}}]{Sasaki2002}%
  \BibitemOpen
  \bibfield  {author} {\bibinfo {author} {\bibfnamefont {T.}~\bibnamefont
  {Sasaki}}, \bibinfo {author} {\bibfnamefont {S.}~\bibnamefont {Sonoda}},
  \bibinfo {author} {\bibfnamefont {Y.}~\bibnamefont {Yamamoto}}, \bibinfo
  {author} {\bibfnamefont {K.~I.}\ \bibnamefont {Suga}}, \bibinfo {author}
  {\bibfnamefont {S.}~\bibnamefont {Shimizu}}, \bibinfo {author} {\bibfnamefont
  {K.}~\bibnamefont {Kindo}}, \ and\ \bibinfo {author} {\bibfnamefont
  {H.}~\bibnamefont {Hori}},\ }\href {\doibase 10.1063/1.1451879} {\bibfield
  {journal} {\bibinfo  {journal} {J. Appl. Phys.}\ }\textbf {\bibinfo {volume}
  {91}},\ \bibinfo {pages} {7911} (\bibinfo {year} {2002})}\BibitemShut
  {NoStop}%
\bibitem [{\citenamefont {Sheu}\ \emph {et~al.}(2009)\citenamefont {Sheu},
  \citenamefont {Yang}, \citenamefont {Tu}, \citenamefont {Chang},
  \citenamefont {Lee}, \citenamefont {Lai},\ and\ \citenamefont
  {Peng}}]{Sheu2009}%
  \BibitemOpen
  \bibfield  {author} {\bibinfo {author} {\bibfnamefont {J.~K.}\ \bibnamefont
  {Sheu}}, \bibinfo {author} {\bibfnamefont {C.~C.}\ \bibnamefont {Yang}},
  \bibinfo {author} {\bibfnamefont {S.~J.}\ \bibnamefont {Tu}}, \bibinfo
  {author} {\bibfnamefont {K.~H.}\ \bibnamefont {Chang}}, \bibinfo {author}
  {\bibfnamefont {M.~L.}\ \bibnamefont {Lee}}, \bibinfo {author} {\bibfnamefont
  {W.~C.}\ \bibnamefont {Lai}}, \ and\ \bibinfo {author} {\bibfnamefont
  {L.~C.}\ \bibnamefont {Peng}},\ }\href {\doibase 10.1109/LED.2008.2012275}
  {\bibfield  {journal} {\bibinfo  {journal} {IEEE Electron Device Lett.}\
  }\textbf {\bibinfo {volume} {30}},\ \bibinfo {pages} {225} (\bibinfo {year}
  {2009})}\BibitemShut {NoStop}%
\bibitem [{\citenamefont {Kuwahara}\ \emph {et~al.}(2010)\citenamefont
  {Kuwahara}, \citenamefont {Fujii}, \citenamefont {Fujiyama}, \citenamefont
  {Sugiyama}, \citenamefont {Iwaya}, \citenamefont {Takeuchi}, \citenamefont
  {Kamiyama}, \citenamefont {Akasaki},\ and\ \citenamefont
  {Amano}}]{Kuwahara2010}%
  \BibitemOpen
  \bibfield  {author} {\bibinfo {author} {\bibfnamefont {Y.}~\bibnamefont
  {Kuwahara}}, \bibinfo {author} {\bibfnamefont {T.}~\bibnamefont {Fujii}},
  \bibinfo {author} {\bibfnamefont {Y.}~\bibnamefont {Fujiyama}}, \bibinfo
  {author} {\bibfnamefont {T.}~\bibnamefont {Sugiyama}}, \bibinfo {author}
  {\bibfnamefont {M.}~\bibnamefont {Iwaya}}, \bibinfo {author} {\bibfnamefont
  {T.}~\bibnamefont {Takeuchi}}, \bibinfo {author} {\bibfnamefont
  {S.}~\bibnamefont {Kamiyama}}, \bibinfo {author} {\bibfnamefont
  {I.}~\bibnamefont {Akasaki}}, \ and\ \bibinfo {author} {\bibfnamefont
  {H.}~\bibnamefont {Amano}},\ }\href {\doibase 10.1143/APEX.3.111001}
  {\bibfield  {journal} {\bibinfo  {journal} {Appl. Phys. Express}\ }\textbf
  {\bibinfo {volume} {3}},\ \bibinfo {pages} {111001} (\bibinfo {year}
  {2010})}\BibitemShut {NoStop}%
\bibitem [{\citenamefont {Wang}\ \emph {et~al.}(2011)\citenamefont {Wang},
  \citenamefont {Pierre}, \citenamefont {Kibria}, \citenamefont {Cui},
  \citenamefont {Han}, \citenamefont {Bevan}, \citenamefont {Guo},
  \citenamefont {Paradis}, \citenamefont {Hakima},\ and\ \citenamefont
  {Mi}}]{Wang2011a}%
  \BibitemOpen
  \bibfield  {author} {\bibinfo {author} {\bibfnamefont {D.}~\bibnamefont
  {Wang}}, \bibinfo {author} {\bibfnamefont {A.}~\bibnamefont {Pierre}},
  \bibinfo {author} {\bibfnamefont {M.~G.}\ \bibnamefont {Kibria}}, \bibinfo
  {author} {\bibfnamefont {K.}~\bibnamefont {Cui}}, \bibinfo {author}
  {\bibfnamefont {X.}~\bibnamefont {Han}}, \bibinfo {author} {\bibfnamefont
  {K.~H.}\ \bibnamefont {Bevan}}, \bibinfo {author} {\bibfnamefont
  {H.}~\bibnamefont {Guo}}, \bibinfo {author} {\bibfnamefont {S.}~\bibnamefont
  {Paradis}}, \bibinfo {author} {\bibfnamefont {A.~R.}\ \bibnamefont {Hakima}},
  \ and\ \bibinfo {author} {\bibfnamefont {Z.}~\bibnamefont {Mi}},\ }\href
  {\doibase 10.1021/nl2006802} {\bibfield  {journal} {\bibinfo  {journal} {Nano
  Lett.}\ }\textbf {\bibinfo {volume} {11}},\ \bibinfo {pages} {2353} (\bibinfo
  {year} {2011})}\BibitemShut {NoStop}%
\bibitem [{\citenamefont {Pearton}\ \emph {et~al.}(2004)\citenamefont
  {Pearton}, \citenamefont {Kang}, \citenamefont {Kim}, \citenamefont {Ren},
  \citenamefont {Gila}, \citenamefont {Abernathy}, \citenamefont {Lin},\ and\
  \citenamefont {Chu}}]{Pearton2004a}%
  \BibitemOpen
  \bibfield  {author} {\bibinfo {author} {\bibfnamefont {S.~J.}\ \bibnamefont
  {Pearton}}, \bibinfo {author} {\bibfnamefont {B.~S.}\ \bibnamefont {Kang}},
  \bibinfo {author} {\bibfnamefont {S.}~\bibnamefont {Kim}}, \bibinfo {author}
  {\bibfnamefont {F.}~\bibnamefont {Ren}}, \bibinfo {author} {\bibfnamefont
  {B.~P.}\ \bibnamefont {Gila}}, \bibinfo {author} {\bibfnamefont {C.~R.}\
  \bibnamefont {Abernathy}}, \bibinfo {author} {\bibfnamefont {J.}~\bibnamefont
  {Lin}}, \ and\ \bibinfo {author} {\bibfnamefont {S.~N.~G.}\ \bibnamefont
  {Chu}},\ }\href {\doibase 10.1088/0953-8984/16/29/R02} {\bibfield  {journal}
  {\bibinfo  {journal} {J. Phys. Condens. Matter}\ }\textbf {\bibinfo {volume}
  {16}},\ \bibinfo {pages} {R961} (\bibinfo {year} {2004})}\BibitemShut
  {NoStop}%
\bibitem [{\citenamefont {Muraro}\ \emph {et~al.}(2010)\citenamefont {Muraro},
  \citenamefont {Nicolas}, \citenamefont {Nhut}, \citenamefont {Forestier},
  \citenamefont {Rochette}, \citenamefont {Vendier}, \citenamefont {Langrez},
  \citenamefont {Cazaux},\ and\ \citenamefont {Feudale}}]{Muraro2010}%
  \BibitemOpen
  \bibfield  {author} {\bibinfo {author} {\bibfnamefont {J.-L.}\ \bibnamefont
  {Muraro}}, \bibinfo {author} {\bibfnamefont {G.}~\bibnamefont {Nicolas}},
  \bibinfo {author} {\bibfnamefont {D.~M.}\ \bibnamefont {Nhut}}, \bibinfo
  {author} {\bibfnamefont {S.}~\bibnamefont {Forestier}}, \bibinfo {author}
  {\bibfnamefont {S.}~\bibnamefont {Rochette}}, \bibinfo {author}
  {\bibfnamefont {O.}~\bibnamefont {Vendier}}, \bibinfo {author} {\bibfnamefont
  {D.}~\bibnamefont {Langrez}}, \bibinfo {author} {\bibfnamefont {J.-L.}\
  \bibnamefont {Cazaux}}, \ and\ \bibinfo {author} {\bibfnamefont
  {M.}~\bibnamefont {Feudale}},\ }\href {\doibase 10.1017/S1759078710000206}
  {\bibfield  {journal} {\bibinfo  {journal} {Int. J. Microw. Wirel. Technol.}\
  }\textbf {\bibinfo {volume} {2}},\ \bibinfo {pages} {121} (\bibinfo {year}
  {2010})}\BibitemShut {NoStop}%
\bibitem [{\citenamefont {Miwa}\ \emph {et~al.}(2011)\citenamefont {Miwa},
  \citenamefont {Kamo}, \citenamefont {Kittaka}, \citenamefont {Yamasaki},
  \citenamefont {Tsukahara}, \citenamefont {Tanii}, \citenamefont {Kohno},
  \citenamefont {Goto},\ and\ \citenamefont {Shima}}]{Miwa2011}%
  \BibitemOpen
  \bibfield  {author} {\bibinfo {author} {\bibfnamefont {S.}~\bibnamefont
  {Miwa}}, \bibinfo {author} {\bibfnamefont {Y.}~\bibnamefont {Kamo}}, \bibinfo
  {author} {\bibfnamefont {Y.}~\bibnamefont {Kittaka}}, \bibinfo {author}
  {\bibfnamefont {T.}~\bibnamefont {Yamasaki}}, \bibinfo {author}
  {\bibfnamefont {Y.}~\bibnamefont {Tsukahara}}, \bibinfo {author}
  {\bibfnamefont {T.}~\bibnamefont {Tanii}}, \bibinfo {author} {\bibfnamefont
  {M.}~\bibnamefont {Kohno}}, \bibinfo {author} {\bibfnamefont
  {S.}~\bibnamefont {Goto}}, \ and\ \bibinfo {author} {\bibfnamefont
  {A.}~\bibnamefont {Shima}},\ }\bibfield  {booktitle} {\emph {\bibinfo
  {booktitle} {Microwave Symposium Digest (MTT), 2011 IEEE MTT-S
  International}},\ }\href {\doibase 10.1109/MWSYM.2011.5972673} {\ ,\ \bibinfo
  {pages} {1} (\bibinfo {year} {2011})}\BibitemShut {NoStop}%
\bibitem [{\citenamefont {Kuykendall}\ \emph {et~al.}(2007)\citenamefont
  {Kuykendall}, \citenamefont {Ulrich}, \citenamefont {Aloni},\ and\
  \citenamefont {Yang}}]{Kuykendall2007}%
  \BibitemOpen
  \bibfield  {author} {\bibinfo {author} {\bibfnamefont {T.}~\bibnamefont
  {Kuykendall}}, \bibinfo {author} {\bibfnamefont {P.}~\bibnamefont {Ulrich}},
  \bibinfo {author} {\bibfnamefont {S.}~\bibnamefont {Aloni}}, \ and\ \bibinfo
  {author} {\bibfnamefont {P.}~\bibnamefont {Yang}},\ }\href {\doibase
  10.1038/nmat2037} {\bibfield  {journal} {\bibinfo  {journal} {Nat. Mater.}\
  }\textbf {\bibinfo {volume} {6}},\ \bibinfo {pages} {951} (\bibinfo {year}
  {2007})}\BibitemShut {NoStop}%
\bibitem [{\citenamefont {Pimputkar}\ \emph {et~al.}(2009)\citenamefont
  {Pimputkar}, \citenamefont {Speck}, \citenamefont {DenBaars},\ and\
  \citenamefont {Nakamura}}]{Pimputkar2009}%
  \BibitemOpen
  \bibfield  {author} {\bibinfo {author} {\bibfnamefont {S.}~\bibnamefont
  {Pimputkar}}, \bibinfo {author} {\bibfnamefont {J.~S.}\ \bibnamefont
  {Speck}}, \bibinfo {author} {\bibfnamefont {S.~P.}\ \bibnamefont {DenBaars}},
  \ and\ \bibinfo {author} {\bibfnamefont {S.}~\bibnamefont {Nakamura}},\
  }\href {\doibase 10.1038/nphoton.2009.32} {\bibfield  {journal} {\bibinfo
  {journal} {Nat. Photonics}\ }\textbf {\bibinfo {volume} {3}},\ \bibinfo
  {pages} {180} (\bibinfo {year} {2009})}\BibitemShut {NoStop}%
\bibitem [{\citenamefont {Wiesmann}\ \emph {et~al.}(2009)\citenamefont
  {Wiesmann}, \citenamefont {Bergenek}, \citenamefont {Linder},\ and\
  \citenamefont {Schwarz}}]{Wiesmann2009a}%
  \BibitemOpen
  \bibfield  {author} {\bibinfo {author} {\bibfnamefont {C.}~\bibnamefont
  {Wiesmann}}, \bibinfo {author} {\bibfnamefont {K.}~\bibnamefont {Bergenek}},
  \bibinfo {author} {\bibfnamefont {N.}~\bibnamefont {Linder}}, \ and\ \bibinfo
  {author} {\bibfnamefont {U.~T.}\ \bibnamefont {Schwarz}},\ }\href {\doibase
  10.1002/lpor.200810053} {\bibfield  {journal} {\bibinfo  {journal} {Laser
  Photonics Rev.}\ }\textbf {\bibinfo {volume} {3}},\ \bibinfo {pages} {262}
  (\bibinfo {year} {2009})}\BibitemShut {NoStop}%
\bibitem [{\citenamefont {Choi}\ \emph {et~al.}(2003)\citenamefont {Choi},
  \citenamefont {Dawson}, \citenamefont {Edwards},\ and\ \citenamefont
  {Martin}}]{Choi2003}%
  \BibitemOpen
  \bibfield  {author} {\bibinfo {author} {\bibfnamefont {H.~W.}\ \bibnamefont
  {Choi}}, \bibinfo {author} {\bibfnamefont {M.~D.}\ \bibnamefont {Dawson}},
  \bibinfo {author} {\bibfnamefont {P.~R.}\ \bibnamefont {Edwards}}, \ and\
  \bibinfo {author} {\bibfnamefont {R.~W.}\ \bibnamefont {Martin}},\ }\href
  {\doibase 10.1063/1.1630352} {\bibfield  {journal} {\bibinfo  {journal}
  {Appl. Phys. Lett.}\ }\textbf {\bibinfo {volume} {83}},\ \bibinfo {pages}
  {4483} (\bibinfo {year} {2003})}\BibitemShut {NoStop}%
\bibitem [{\citenamefont {David}, \citenamefont {Benisty},\ and\ \citenamefont
  {Weisbuch}(2007)}]{David2007}%
  \BibitemOpen
  \bibfield  {author} {\bibinfo {author} {\bibfnamefont {A.}~\bibnamefont
  {David}}, \bibinfo {author} {\bibfnamefont {H.}~\bibnamefont {Benisty}}, \
  and\ \bibinfo {author} {\bibfnamefont {C.}~\bibnamefont {Weisbuch}},\ }\href
  {\doibase 10.1109/JDT.2007.896736} {\bibfield  {journal} {\bibinfo  {journal}
  {J. Disp. Technol.}\ }\textbf {\bibinfo {volume} {3}},\ \bibinfo {pages}
  {133} (\bibinfo {year} {2007})}\BibitemShut {NoStop}%
\bibitem [{\citenamefont {Bilousov}\ \emph {et~al.}(2014)\citenamefont
  {Bilousov}, \citenamefont {Carvajal}, \citenamefont {Geaney}, \citenamefont
  {Zubialevich}, \citenamefont {Parbrook}, \citenamefont {Mart}, \citenamefont
  {Jime}, \citenamefont {Francesc}, \citenamefont {Aguilo},\ and\ \citenamefont
  {Dwyer}}]{Bilousov2014}%
  \BibitemOpen
  \bibfield  {author} {\bibinfo {author} {\bibfnamefont {O.~V.}\ \bibnamefont
  {Bilousov}}, \bibinfo {author} {\bibfnamefont {J.~J.}\ \bibnamefont
  {Carvajal}}, \bibinfo {author} {\bibfnamefont {H.}~\bibnamefont {Geaney}},
  \bibinfo {author} {\bibfnamefont {V.~Z.}\ \bibnamefont {Zubialevich}},
  \bibinfo {author} {\bibfnamefont {P.~J.}\ \bibnamefont {Parbrook}}, \bibinfo
  {author} {\bibfnamefont {O.}~\bibnamefont {Mart}}, \bibinfo {author}
  {\bibfnamefont {J.}~\bibnamefont {Jime}}, \bibinfo {author} {\bibfnamefont
  {D.}~\bibnamefont {Francesc}}, \bibinfo {author} {\bibfnamefont
  {M.}~\bibnamefont {Aguilo}}, \ and\ \bibinfo {author} {\bibfnamefont {C.~O.}\
  \bibnamefont {Dwyer}},\ }\href@noop {} {\bibfield  {journal} {\bibinfo
  {journal} {ACS Appl. Mater. Interfaces}\ }\textbf {\bibinfo {volume} {6}},\
  \bibinfo {pages} {17954} (\bibinfo {year} {2014})}\BibitemShut {NoStop}%
\bibitem [{\citenamefont {Soh}\ \emph {et~al.}(2013)\citenamefont {Soh},
  \citenamefont {Tay}, \citenamefont {Tan}, \citenamefont {Vajpeyi},
  \citenamefont {Seetoh}, \citenamefont {Ansah-Antwi},\ and\ \citenamefont
  {Chua}}]{Soh2013}%
  \BibitemOpen
  \bibfield  {author} {\bibinfo {author} {\bibfnamefont {C.~B.}\ \bibnamefont
  {Soh}}, \bibinfo {author} {\bibfnamefont {C.~B.}\ \bibnamefont {Tay}},
  \bibinfo {author} {\bibfnamefont {R.~J.~N.}\ \bibnamefont {Tan}}, \bibinfo
  {author} {\bibfnamefont {A.~P.}\ \bibnamefont {Vajpeyi}}, \bibinfo {author}
  {\bibfnamefont {I.~P.}\ \bibnamefont {Seetoh}}, \bibinfo {author}
  {\bibfnamefont {K.~K.}\ \bibnamefont {Ansah-Antwi}}, \ and\ \bibinfo {author}
  {\bibfnamefont {S.~J.}\ \bibnamefont {Chua}},\ }\href {\doibase
  10.1088/0022-3727/46/36/365102} {\bibfield  {journal} {\bibinfo  {journal}
  {J. Phys. D. Appl. Phys.}\ }\textbf {\bibinfo {volume} {46}},\ \bibinfo
  {pages} {365102} (\bibinfo {year} {2013})}\BibitemShut {NoStop}%
\bibitem [{\citenamefont {Yang}\ \emph {et~al.}(2008)\citenamefont {Yang},
  \citenamefont {Lin}, \citenamefont {Lin}, \citenamefont {Chang},
  \citenamefont {Chen}, \citenamefont {Chien},\ and\ \citenamefont
  {Chang}}]{Yang2008}%
  \BibitemOpen
  \bibfield  {author} {\bibinfo {author} {\bibfnamefont {C.~C.}\ \bibnamefont
  {Yang}}, \bibinfo {author} {\bibfnamefont {C.~F.}\ \bibnamefont {Lin}},
  \bibinfo {author} {\bibfnamefont {C.~M.}\ \bibnamefont {Lin}}, \bibinfo
  {author} {\bibfnamefont {C.~C.}\ \bibnamefont {Chang}}, \bibinfo {author}
  {\bibfnamefont {K.~T.}\ \bibnamefont {Chen}}, \bibinfo {author}
  {\bibfnamefont {J.~F.}\ \bibnamefont {Chien}}, \ and\ \bibinfo {author}
  {\bibfnamefont {C.~Y.}\ \bibnamefont {Chang}},\ }\href {\doibase
  10.1063/1.3027068} {\bibfield  {journal} {\bibinfo  {journal} {Appl. Phys.
  Lett.}\ }\textbf {\bibinfo {volume} {93}},\ \bibinfo {pages} {2006} (\bibinfo
  {year} {2008})}\BibitemShut {NoStop}%
\bibitem [{\citenamefont {Kesaria}\ \emph {et~al.}(2011)\citenamefont
  {Kesaria}, \citenamefont {Shetty}, \citenamefont {Cohen},\ and\ \citenamefont
  {Shivaprasad}}]{Kesaria2011}%
  \BibitemOpen
  \bibfield  {author} {\bibinfo {author} {\bibfnamefont {M.}~\bibnamefont
  {Kesaria}}, \bibinfo {author} {\bibfnamefont {S.}~\bibnamefont {Shetty}},
  \bibinfo {author} {\bibfnamefont {P.}~\bibnamefont {Cohen}}, \ and\ \bibinfo
  {author} {\bibfnamefont {S.}~\bibnamefont {Shivaprasad}},\ }\href {\doibase
  10.1016/j.materresbull.2011.07.043} {\bibfield  {journal} {\bibinfo
  {journal} {Mater. Res. Bull.}\ }\textbf {\bibinfo {volume} {46}},\ \bibinfo
  {pages} {1811} (\bibinfo {year} {2011})}\BibitemShut {NoStop}%
\bibitem [{\citenamefont {Thakur}\ \emph {et~al.}(2015)\citenamefont {Thakur},
  \citenamefont {Nayak}, \citenamefont {Nagaraja},\ and\ \citenamefont
  {Shivaprasad}}]{Thakur2015}%
  \BibitemOpen
  \bibfield  {author} {\bibinfo {author} {\bibfnamefont {V.}~\bibnamefont
  {Thakur}}, \bibinfo {author} {\bibfnamefont {S.~K.}\ \bibnamefont {Nayak}},
  \bibinfo {author} {\bibfnamefont {K.~K.}\ \bibnamefont {Nagaraja}}, \ and\
  \bibinfo {author} {\bibfnamefont {S.~M.}\ \bibnamefont {Shivaprasad}},\
  }\href {\doibase 10.1007/s13391-015-4388-3} {\bibfield  {journal} {\bibinfo
  {journal} {Electron. Mater. Lett.}\ }\textbf {\bibinfo {volume} {11}},\
  \bibinfo {pages} {398} (\bibinfo {year} {2015})}\BibitemShut {NoStop}%
\bibitem [{\citenamefont {Nayak}\ \emph {et~al.}(2016)\citenamefont {Nayak},
  \citenamefont {Shamoon}, \citenamefont {Ghatak},\ and\ \citenamefont
  {Shivaprasad}}]{Nayak2016a}%
  \BibitemOpen
  \bibfield  {author} {\bibinfo {author} {\bibfnamefont {S.~K.}\ \bibnamefont
  {Nayak}}, \bibinfo {author} {\bibfnamefont {D.}~\bibnamefont {Shamoon}},
  \bibinfo {author} {\bibfnamefont {J.}~\bibnamefont {Ghatak}}, \ and\ \bibinfo
  {author} {\bibfnamefont {S.}~\bibnamefont {Shivaprasad}},\ }\href {\doibase
  10.1002/pssa.201600300} {\bibfield  {journal} {\bibinfo  {journal} {Phys.
  Status Solidi}\ } (\bibinfo {year} {2016}),\
  10.1002/pssa.201600300}\BibitemShut {NoStop}%
\bibitem [{\citenamefont {Zhang}\ \emph {et~al.}(2010)\citenamefont {Zhang},
  \citenamefont {Bhattacharya}, \citenamefont {Guo},\ and\ \citenamefont
  {Banerjee}}]{Zhang2010}%
  \BibitemOpen
  \bibfield  {author} {\bibinfo {author} {\bibfnamefont {M.}~\bibnamefont
  {Zhang}}, \bibinfo {author} {\bibfnamefont {P.}~\bibnamefont {Bhattacharya}},
  \bibinfo {author} {\bibfnamefont {W.}~\bibnamefont {Guo}}, \ and\ \bibinfo
  {author} {\bibfnamefont {A.}~\bibnamefont {Banerjee}},\ }\href {\doibase
  10.1063/1.3374882} {\bibfield  {journal} {\bibinfo  {journal} {Appl. Phys.
  Lett.}\ }\textbf {\bibinfo {volume} {96}},\ \bibinfo {pages} {132103}
  (\bibinfo {year} {2010})}\BibitemShut {NoStop}%
\bibitem [{\citenamefont {Cimpoiasu}\ \emph {et~al.}(2006)\citenamefont
  {Cimpoiasu}, \citenamefont {Stern}, \citenamefont {Klie}, \citenamefont
  {Munden}, \citenamefont {Cheng},\ and\ \citenamefont {Reed}}]{Cimpoiasu2006}%
  \BibitemOpen
  \bibfield  {author} {\bibinfo {author} {\bibfnamefont {E.}~\bibnamefont
  {Cimpoiasu}}, \bibinfo {author} {\bibfnamefont {E.}~\bibnamefont {Stern}},
  \bibinfo {author} {\bibfnamefont {R.}~\bibnamefont {Klie}}, \bibinfo {author}
  {\bibfnamefont {R.~a.}\ \bibnamefont {Munden}}, \bibinfo {author}
  {\bibfnamefont {G.}~\bibnamefont {Cheng}}, \ and\ \bibinfo {author}
  {\bibfnamefont {M.~a.}\ \bibnamefont {Reed}},\ }\href {\doibase
  10.1088/0957-4484/17/23/004} {\bibfield  {journal} {\bibinfo  {journal}
  {Nanotechnology}\ }\textbf {\bibinfo {volume} {17}},\ \bibinfo {pages} {5735}
  (\bibinfo {year} {2006})}\BibitemShut {NoStop}%
\bibitem [{\citenamefont {Ptak}\ \emph {et~al.}(2001)\citenamefont {Ptak},
  \citenamefont {Holbert}, \citenamefont {Ting}, \citenamefont {Swartz},
  \citenamefont {Moldovan}, \citenamefont {Giles}, \citenamefont {Myers},
  \citenamefont {{Van Lierde}}, \citenamefont {Tian}, \citenamefont {Hockett},
  \citenamefont {Mitha}, \citenamefont {Wickenden}, \citenamefont {Koleske},\
  and\ \citenamefont {Henry}}]{Ptak2001}%
  \BibitemOpen
  \bibfield  {author} {\bibinfo {author} {\bibfnamefont {A.~J.}\ \bibnamefont
  {Ptak}}, \bibinfo {author} {\bibfnamefont {L.~J.}\ \bibnamefont {Holbert}},
  \bibinfo {author} {\bibfnamefont {L.}~\bibnamefont {Ting}}, \bibinfo {author}
  {\bibfnamefont {C.~H.}\ \bibnamefont {Swartz}}, \bibinfo {author}
  {\bibfnamefont {M.}~\bibnamefont {Moldovan}}, \bibinfo {author}
  {\bibfnamefont {N.~C.}\ \bibnamefont {Giles}}, \bibinfo {author}
  {\bibfnamefont {T.~H.}\ \bibnamefont {Myers}}, \bibinfo {author}
  {\bibfnamefont {P.}~\bibnamefont {{Van Lierde}}}, \bibinfo {author}
  {\bibfnamefont {C.}~\bibnamefont {Tian}}, \bibinfo {author} {\bibfnamefont
  {R.~A.}\ \bibnamefont {Hockett}}, \bibinfo {author} {\bibfnamefont
  {S.}~\bibnamefont {Mitha}}, \bibinfo {author} {\bibfnamefont {A.~E.}\
  \bibnamefont {Wickenden}}, \bibinfo {author} {\bibfnamefont {D.~D.}\
  \bibnamefont {Koleske}}, \ and\ \bibinfo {author} {\bibfnamefont {R.~L.}\
  \bibnamefont {Henry}},\ }\href {\doibase 10.1063/1.1403276} {\bibfield
  {journal} {\bibinfo  {journal} {Appl. Phys. Lett.}\ }\textbf {\bibinfo
  {volume} {79}},\ \bibinfo {pages} {2740} (\bibinfo {year}
  {2001})}\BibitemShut {NoStop}%
\bibitem [{\citenamefont {Myers}\ \emph {et~al.}(2001)\citenamefont {Myers},
  \citenamefont {Wright}, \citenamefont {Petersen}, \citenamefont {Wampler},
  \citenamefont {Seager}, \citenamefont {Crawford},\ and\ \citenamefont
  {Han}}]{Myers2001}%
  \BibitemOpen
  \bibfield  {author} {\bibinfo {author} {\bibfnamefont {S.~M.}\ \bibnamefont
  {Myers}}, \bibinfo {author} {\bibfnamefont {A.~F.}\ \bibnamefont {Wright}},
  \bibinfo {author} {\bibfnamefont {G.~A.}\ \bibnamefont {Petersen}}, \bibinfo
  {author} {\bibfnamefont {W.~R.}\ \bibnamefont {Wampler}}, \bibinfo {author}
  {\bibfnamefont {C.~H.}\ \bibnamefont {Seager}}, \bibinfo {author}
  {\bibfnamefont {M.~H.}\ \bibnamefont {Crawford}}, \ and\ \bibinfo {author}
  {\bibfnamefont {J.}~\bibnamefont {Han}},\ }\href {\doibase 10.1063/1.1347410}
  {\bibfield  {journal} {\bibinfo  {journal} {J. Appl. Phys.}\ }\textbf
  {\bibinfo {volume} {89}},\ \bibinfo {pages} {3195} (\bibinfo {year}
  {2001})}\BibitemShut {NoStop}%
\bibitem [{\citenamefont {Smorchkova}\ \emph {et~al.}(2000)\citenamefont
  {Smorchkova}, \citenamefont {Haus}, \citenamefont {Heying}, \citenamefont
  {Kozodoy}, \citenamefont {Fini}, \citenamefont {Ibbetson}, \citenamefont
  {Keller}, \citenamefont {DenBaars}, \citenamefont {Speck},\ and\
  \citenamefont {Mishra}}]{Smorchkova2000}%
  \BibitemOpen
  \bibfield  {author} {\bibinfo {author} {\bibfnamefont {I.~P.}\ \bibnamefont
  {Smorchkova}}, \bibinfo {author} {\bibfnamefont {E.}~\bibnamefont {Haus}},
  \bibinfo {author} {\bibfnamefont {B.}~\bibnamefont {Heying}}, \bibinfo
  {author} {\bibfnamefont {P.}~\bibnamefont {Kozodoy}}, \bibinfo {author}
  {\bibfnamefont {P.}~\bibnamefont {Fini}}, \bibinfo {author} {\bibfnamefont
  {J.~P.}\ \bibnamefont {Ibbetson}}, \bibinfo {author} {\bibfnamefont
  {S.}~\bibnamefont {Keller}}, \bibinfo {author} {\bibfnamefont {S.~P.}\
  \bibnamefont {DenBaars}}, \bibinfo {author} {\bibfnamefont {J.~S.}\
  \bibnamefont {Speck}}, \ and\ \bibinfo {author} {\bibfnamefont {U.~K.}\
  \bibnamefont {Mishra}},\ }\href {\doibase 10.1063/1.125872} {\bibfield
  {journal} {\bibinfo  {journal} {Appl. Phys. Lett.}\ }\textbf {\bibinfo
  {volume} {76}},\ \bibinfo {pages} {718} (\bibinfo {year} {2000})}\BibitemShut
  {NoStop}%
\bibitem [{\citenamefont {Miceli}\ and\ \citenamefont
  {Pasquarello}(2016)}]{Miceli2016}%
  \BibitemOpen
  \bibfield  {author} {\bibinfo {author} {\bibfnamefont {G.}~\bibnamefont
  {Miceli}}\ and\ \bibinfo {author} {\bibfnamefont {A.}~\bibnamefont
  {Pasquarello}},\ }\href {\doibase 10.1103/PhysRevB.93.165207} {\bibfield
  {journal} {\bibinfo  {journal} {Phys. Rev. B}\ }\textbf {\bibinfo {volume}
  {93}},\ \bibinfo {pages} {165207} (\bibinfo {year} {2016})}\BibitemShut
  {NoStop}%
\bibitem [{\citenamefont {Nakano}\ and\ \citenamefont
  {Jimbo}(2002)}]{Nakano2002}%
  \BibitemOpen
  \bibfield  {author} {\bibinfo {author} {\bibfnamefont {Y.}~\bibnamefont
  {Nakano}}\ and\ \bibinfo {author} {\bibfnamefont {T.}~\bibnamefont {Jimbo}},\
  }\href {\doibase 10.1002/pssc.200390082} {\bibfield  {journal} {\bibinfo
  {journal} {Phys. Status Solidi (C)}\ }\textbf {\bibinfo {volume} {442}},\
  \bibinfo {pages} {438} (\bibinfo {year} {2002})}\BibitemShut {NoStop}%
\bibitem [{\citenamefont {Hashizume}(2003)}]{Hashizume2003}%
  \BibitemOpen
  \bibfield  {author} {\bibinfo {author} {\bibfnamefont {T.}~\bibnamefont
  {Hashizume}},\ }\href {\doibase 10.1063/1.1580195} {\bibfield  {journal}
  {\bibinfo  {journal} {J. Appl. Phys.}\ }\textbf {\bibinfo {volume} {94}},\
  \bibinfo {pages} {431} (\bibinfo {year} {2003})}\BibitemShut {NoStop}%
\bibitem [{\citenamefont {Cheng}\ \emph {et~al.}(1999)\citenamefont {Cheng},
  \citenamefont {Novikov}, \citenamefont {Foxon},\ and\ \citenamefont
  {Orton}}]{Cheng1999}%
  \BibitemOpen
  \bibfield  {author} {\bibinfo {author} {\bibfnamefont {T.}~\bibnamefont
  {Cheng}}, \bibinfo {author} {\bibfnamefont {S.}~\bibnamefont {Novikov}},
  \bibinfo {author} {\bibfnamefont {C.}~\bibnamefont {Foxon}}, \ and\ \bibinfo
  {author} {\bibfnamefont {J.}~\bibnamefont {Orton}},\ }\href {\doibase
  10.1016/S0038-1098(98)00601-2} {\bibfield  {journal} {\bibinfo  {journal}
  {Solid State Commun.}\ }\textbf {\bibinfo {volume} {109}},\ \bibinfo {pages}
  {439} (\bibinfo {year} {1999})}\BibitemShut {NoStop}%
\bibitem [{\citenamefont {Romano}\ \emph {et~al.}(2001)\citenamefont {Romano},
  \citenamefont {Kneissl}, \citenamefont {Northrup}, \citenamefont {{Van De
  Walle}},\ and\ \citenamefont {Treat}}]{Romano2001}%
  \BibitemOpen
  \bibfield  {author} {\bibinfo {author} {\bibfnamefont {L.~T.}\ \bibnamefont
  {Romano}}, \bibinfo {author} {\bibfnamefont {M.}~\bibnamefont {Kneissl}},
  \bibinfo {author} {\bibfnamefont {J.~E.}\ \bibnamefont {Northrup}}, \bibinfo
  {author} {\bibfnamefont {C.~G.}\ \bibnamefont {{Van De Walle}}}, \ and\
  \bibinfo {author} {\bibfnamefont {D.~W.}\ \bibnamefont {Treat}},\ }\href
  {\doibase 10.1063/1.1413222} {\bibfield  {journal} {\bibinfo  {journal}
  {Appl. Phys. Lett.}\ }\textbf {\bibinfo {volume} {79}},\ \bibinfo {pages}
  {2734} (\bibinfo {year} {2001})}\BibitemShut {NoStop}%
\bibitem [{\citenamefont {Kesaria}\ and\ \citenamefont
  {Shivaprasad}(2011)}]{Kesaria2011f}%
  \BibitemOpen
  \bibfield  {author} {\bibinfo {author} {\bibfnamefont {M.}~\bibnamefont
  {Kesaria}}\ and\ \bibinfo {author} {\bibfnamefont {S.~M.}\ \bibnamefont
  {Shivaprasad}},\ }\href {\doibase 10.1063/1.3646391} {\bibfield  {journal}
  {\bibinfo  {journal} {Appl. Phys. Lett.}\ }\textbf {\bibinfo {volume} {99}},\
  \bibinfo {pages} {143105} (\bibinfo {year} {2011})}\BibitemShut {NoStop}%
\bibitem [{\citenamefont {Reshchikov}\ \emph {et~al.}(2014)\citenamefont
  {Reshchikov}, \citenamefont {Demchenko}, \citenamefont {McNamara},
  \citenamefont {Fern{\'{a}}ndez-Garrido},\ and\ \citenamefont
  {Calarco}}]{Reshchikov2014}%
  \BibitemOpen
  \bibfield  {author} {\bibinfo {author} {\bibfnamefont {M.~A.}\ \bibnamefont
  {Reshchikov}}, \bibinfo {author} {\bibfnamefont {D.~O.}\ \bibnamefont
  {Demchenko}}, \bibinfo {author} {\bibfnamefont {J.~D.}\ \bibnamefont
  {McNamara}}, \bibinfo {author} {\bibfnamefont {S.}~\bibnamefont
  {Fern{\'{a}}ndez-Garrido}}, \ and\ \bibinfo {author} {\bibfnamefont
  {R.}~\bibnamefont {Calarco}},\ }\href {\doibase 10.1103/PhysRevB.90.035207}
  {\bibfield  {journal} {\bibinfo  {journal} {Phys. Rev. B}\ }\textbf {\bibinfo
  {volume} {90}},\ \bibinfo {pages} {035207} (\bibinfo {year}
  {2014})}\BibitemShut {NoStop}%
\bibitem [{\citenamefont {Kirste}\ \emph {et~al.}(2013)\citenamefont {Kirste},
  \citenamefont {Hoffmann}, \citenamefont {Tweedie}, \citenamefont {Bryan},
  \citenamefont {Callsen}, \citenamefont {Kure}, \citenamefont {Nenstiel},
  \citenamefont {Wagner}, \citenamefont {Collazo}, \citenamefont {Hoffmann},\
  and\ \citenamefont {Sitar}}]{Kirste2013}%
  \BibitemOpen
  \bibfield  {author} {\bibinfo {author} {\bibfnamefont {R.}~\bibnamefont
  {Kirste}}, \bibinfo {author} {\bibfnamefont {M.~P.}\ \bibnamefont
  {Hoffmann}}, \bibinfo {author} {\bibfnamefont {J.}~\bibnamefont {Tweedie}},
  \bibinfo {author} {\bibfnamefont {Z.}~\bibnamefont {Bryan}}, \bibinfo
  {author} {\bibfnamefont {G.}~\bibnamefont {Callsen}}, \bibinfo {author}
  {\bibfnamefont {T.}~\bibnamefont {Kure}}, \bibinfo {author} {\bibfnamefont
  {C.}~\bibnamefont {Nenstiel}}, \bibinfo {author} {\bibfnamefont {M.~R.}\
  \bibnamefont {Wagner}}, \bibinfo {author} {\bibfnamefont {R.}~\bibnamefont
  {Collazo}}, \bibinfo {author} {\bibfnamefont {A.}~\bibnamefont {Hoffmann}}, \
  and\ \bibinfo {author} {\bibfnamefont {Z.}~\bibnamefont {Sitar}},\ }\href
  {\doibase 10.1063/1.4794094} {\bibfield  {journal} {\bibinfo  {journal} {J.
  Appl. Phys.}\ }\textbf {\bibinfo {volume} {113}},\ \bibinfo {pages} {103504}
  (\bibinfo {year} {2013})}\BibitemShut {NoStop}%
\bibitem [{\citenamefont {Gelhausen}\ \emph {et~al.}(2004)\citenamefont
  {Gelhausen}, \citenamefont {Phillips}, \citenamefont {Goldys}, \citenamefont
  {Paskova}, \citenamefont {Monemar}, \citenamefont {Strassburg},\ and\
  \citenamefont {Hoffmann}}]{Gelhausen2004}%
  \BibitemOpen
  \bibfield  {author} {\bibinfo {author} {\bibfnamefont {O.}~\bibnamefont
  {Gelhausen}}, \bibinfo {author} {\bibfnamefont {M.}~\bibnamefont {Phillips}},
  \bibinfo {author} {\bibfnamefont {E.}~\bibnamefont {Goldys}}, \bibinfo
  {author} {\bibfnamefont {T.}~\bibnamefont {Paskova}}, \bibinfo {author}
  {\bibfnamefont {B.}~\bibnamefont {Monemar}}, \bibinfo {author} {\bibfnamefont
  {M.}~\bibnamefont {Strassburg}}, \ and\ \bibinfo {author} {\bibfnamefont
  {A.}~\bibnamefont {Hoffmann}},\ }\href {\doibase 10.1103/PhysRevB.69.125210}
  {\bibfield  {journal} {\bibinfo  {journal} {Phys. Rev. B}\ }\textbf {\bibinfo
  {volume} {69}},\ \bibinfo {pages} {125210} (\bibinfo {year}
  {2004})}\BibitemShut {NoStop}%
\bibitem [{\citenamefont {Harima}(2002)}]{Harima2002}%
  \BibitemOpen
  \bibfield  {author} {\bibinfo {author} {\bibfnamefont {H.}~\bibnamefont
  {Harima}},\ }\href {\doibase 10.1088/0953-8984/14/38/201} {\bibfield
  {journal} {\bibinfo  {journal} {J. Phys. Condens. Matter}\ }\textbf {\bibinfo
  {volume} {14}},\ \bibinfo {pages} {R967} (\bibinfo {year}
  {2002})}\BibitemShut {NoStop}%
\bibitem [{\citenamefont {Zhang}\ \emph {et~al.}(2013)\citenamefont {Zhang},
  \citenamefont {Liu}, \citenamefont {Si}, \citenamefont {Ren}, \citenamefont
  {Wang}, \citenamefont {Feng}, \citenamefont {Dong}, \citenamefont {Du},
  \citenamefont {Zhu}, \citenamefont {Fu}, \citenamefont {Lu}, \citenamefont
  {Li},\ and\ \citenamefont {Wang}}]{Zhang2013a}%
  \BibitemOpen
  \bibfield  {author} {\bibinfo {author} {\bibfnamefont {N.}~\bibnamefont
  {Zhang}}, \bibinfo {author} {\bibfnamefont {Z.}~\bibnamefont {Liu}}, \bibinfo
  {author} {\bibfnamefont {Z.}~\bibnamefont {Si}}, \bibinfo {author}
  {\bibfnamefont {P.}~\bibnamefont {Ren}}, \bibinfo {author} {\bibfnamefont
  {X.-D.}\ \bibnamefont {Wang}}, \bibinfo {author} {\bibfnamefont {X.-X.}\
  \bibnamefont {Feng}}, \bibinfo {author} {\bibfnamefont {P.}~\bibnamefont
  {Dong}}, \bibinfo {author} {\bibfnamefont {C.-X.}\ \bibnamefont {Du}},
  \bibinfo {author} {\bibfnamefont {S.-X.}\ \bibnamefont {Zhu}}, \bibinfo
  {author} {\bibfnamefont {B.-L.}\ \bibnamefont {Fu}}, \bibinfo {author}
  {\bibfnamefont {H.-X.}\ \bibnamefont {Lu}}, \bibinfo {author} {\bibfnamefont
  {J.-M.}\ \bibnamefont {Li}}, \ and\ \bibinfo {author} {\bibfnamefont {J.-X.}\
  \bibnamefont {Wang}},\ }\href {\doibase 10.1088/0256-307X/30/8/087101}
  {\bibfield  {journal} {\bibinfo  {journal} {Chinese Phys. Lett.}\ }\textbf
  {\bibinfo {volume} {30}},\ \bibinfo {pages} {087101} (\bibinfo {year}
  {2013})}\BibitemShut {NoStop}%
\bibitem [{\citenamefont {Reshchikov}, \citenamefont {Yi},\ and\ \citenamefont
  {Wessels}(1999)}]{Reshchikov1999}%
  \BibitemOpen
  \bibfield  {author} {\bibinfo {author} {\bibfnamefont {M.}~\bibnamefont
  {Reshchikov}}, \bibinfo {author} {\bibfnamefont {G.-C.}\ \bibnamefont {Yi}},
  \ and\ \bibinfo {author} {\bibfnamefont {B.}~\bibnamefont {Wessels}},\ }\href
  {\doibase 10.1103/PhysRevB.59.13176} {\bibfield  {journal} {\bibinfo
  {journal} {Phys. Rev. B}\ }\textbf {\bibinfo {volume} {59}},\ \bibinfo
  {pages} {13176} (\bibinfo {year} {1999})}\BibitemShut {NoStop}%
\bibitem [{\citenamefont {Monemar}\ \emph {et~al.}(2014)\citenamefont
  {Monemar}, \citenamefont {Paskov}, \citenamefont {Pozina}, \citenamefont
  {Hemmingsson}, \citenamefont {Bergman}, \citenamefont {Khromov},
  \citenamefont {Izyumskaya}, \citenamefont {Avrutin}, \citenamefont {Li},
  \citenamefont {Morkoc}, \citenamefont {Amano}, \citenamefont {Iwaya},\ and\
  \citenamefont {Akasaki}}]{Monemar2014}%
  \BibitemOpen
  \bibfield  {author} {\bibinfo {author} {\bibfnamefont {B.}~\bibnamefont
  {Monemar}}, \bibinfo {author} {\bibfnamefont {P.~P.}\ \bibnamefont {Paskov}},
  \bibinfo {author} {\bibfnamefont {G.}~\bibnamefont {Pozina}}, \bibinfo
  {author} {\bibfnamefont {C.}~\bibnamefont {Hemmingsson}}, \bibinfo {author}
  {\bibfnamefont {J.~P.}\ \bibnamefont {Bergman}}, \bibinfo {author}
  {\bibfnamefont {S.}~\bibnamefont {Khromov}}, \bibinfo {author} {\bibfnamefont
  {V.~N.}\ \bibnamefont {Izyumskaya}}, \bibinfo {author} {\bibfnamefont
  {V.}~\bibnamefont {Avrutin}}, \bibinfo {author} {\bibfnamefont
  {X.}~\bibnamefont {Li}}, \bibinfo {author} {\bibfnamefont {H.}~\bibnamefont
  {Morkoc}}, \bibinfo {author} {\bibfnamefont {H.}~\bibnamefont {Amano}},
  \bibinfo {author} {\bibfnamefont {M.}~\bibnamefont {Iwaya}}, \ and\ \bibinfo
  {author} {\bibfnamefont {I.}~\bibnamefont {Akasaki}},\ }\href {\doibase
  10.1063/1.4862928} {\bibfield  {journal} {\bibinfo  {journal} {J. Appl.
  Phys.}\ }\textbf {\bibinfo {volume} {115}},\ \bibinfo {pages} {053507}
  (\bibinfo {year} {2014})}\BibitemShut {NoStop}%
\bibitem [{\citenamefont {Kaufmann}\ \emph {et~al.}(1998)\citenamefont
  {Kaufmann}, \citenamefont {Kunzer}, \citenamefont {Maier}, \citenamefont
  {Obloh}, \citenamefont {Ramakrishnan}, \citenamefont {Santic},\ and\
  \citenamefont {Schlotter}}]{kaufmann1998}%
  \BibitemOpen
  \bibfield  {author} {\bibinfo {author} {\bibfnamefont {U.}~\bibnamefont
  {Kaufmann}}, \bibinfo {author} {\bibfnamefont {M.}~\bibnamefont {Kunzer}},
  \bibinfo {author} {\bibfnamefont {M.}~\bibnamefont {Maier}}, \bibinfo
  {author} {\bibfnamefont {H.}~\bibnamefont {Obloh}}, \bibinfo {author}
  {\bibfnamefont {A.}~\bibnamefont {Ramakrishnan}}, \bibinfo {author}
  {\bibfnamefont {B.}~\bibnamefont {Santic}}, \ and\ \bibinfo {author}
  {\bibfnamefont {P.}~\bibnamefont {Schlotter}},\ }\href@noop {} {\bibfield
  {journal} {\bibinfo  {journal} {Appl. Phys. Lett.}\ }\textbf {\bibinfo
  {volume} {72}},\ \bibinfo {pages} {1326} (\bibinfo {year}
  {1998})}\BibitemShut {NoStop}%
\bibitem [{\citenamefont {Yan}\ \emph {et~al.}(2012)\citenamefont {Yan},
  \citenamefont {Janotti}, \citenamefont {Scheffler}, \citenamefont {Walle},
  \citenamefont {Yan}, \citenamefont {Janotti}, \citenamefont {Scheffler},\
  and\ \citenamefont {Walle}}]{Qimin2012}%
  \BibitemOpen
  \bibfield  {author} {\bibinfo {author} {\bibfnamefont {Q.}~\bibnamefont
  {Yan}}, \bibinfo {author} {\bibfnamefont {A.}~\bibnamefont {Janotti}},
  \bibinfo {author} {\bibfnamefont {M.}~\bibnamefont {Scheffler}}, \bibinfo
  {author} {\bibfnamefont {C.~G. V.~D.}\ \bibnamefont {Walle}}, \bibinfo
  {author} {\bibfnamefont {Q.}~\bibnamefont {Yan}}, \bibinfo {author}
  {\bibfnamefont {A.}~\bibnamefont {Janotti}}, \bibinfo {author} {\bibfnamefont
  {M.}~\bibnamefont {Scheffler}}, \ and\ \bibinfo {author} {\bibfnamefont
  {C.~G. V.~D.}\ \bibnamefont {Walle}},\ }\href {\doibase 10.1063/1.3699009}
  {\bibfield  {journal} {\bibinfo  {journal} {Appl. Phys. Lett.}\ }\textbf
  {\bibinfo {volume} {100}},\ \bibinfo {pages} {142110} (\bibinfo {year}
  {2012})}\BibitemShut {NoStop}%
\bibitem [{\citenamefont {Buckeridge}\ \emph {et~al.}(2015)\citenamefont
  {Buckeridge}, \citenamefont {Catlow}, \citenamefont {Scanlon}, \citenamefont
  {Keal}, \citenamefont {Sherwood}, \citenamefont {Miskufova}, \citenamefont
  {Walsh}, \citenamefont {Woodley},\ and\ \citenamefont
  {Sokol}}]{Buckeridge2015}%
  \BibitemOpen
  \bibfield  {author} {\bibinfo {author} {\bibfnamefont {J.}~\bibnamefont
  {Buckeridge}}, \bibinfo {author} {\bibfnamefont {C.~R.~A.}\ \bibnamefont
  {Catlow}}, \bibinfo {author} {\bibfnamefont {D.~O.}\ \bibnamefont {Scanlon}},
  \bibinfo {author} {\bibfnamefont {T.}~\bibnamefont {Keal}}, \bibinfo {author}
  {\bibfnamefont {P.}~\bibnamefont {Sherwood}}, \bibinfo {author}
  {\bibfnamefont {M.}~\bibnamefont {Miskufova}}, \bibinfo {author}
  {\bibfnamefont {A.}~\bibnamefont {Walsh}}, \bibinfo {author} {\bibfnamefont
  {S.~M.}\ \bibnamefont {Woodley}}, \ and\ \bibinfo {author} {\bibfnamefont
  {A.~A.}\ \bibnamefont {Sokol}},\ }\href@noop {} {\bibfield  {journal}
  {\bibinfo  {journal} {Phys. Rev. Lett.}\ }\textbf {\bibinfo {volume} {114}},\
  \bibinfo {pages} {016405} (\bibinfo {year} {2015})}\BibitemShut {NoStop}%
\bibitem [{\citenamefont {Kaufmann}\ \emph {et~al.}(1999)\citenamefont
  {Kaufmann}, \citenamefont {Kunzer}, \citenamefont {Obloh}, \citenamefont
  {Maier}, \citenamefont {Manz}, \citenamefont {Ramakrishnan},\ and\
  \citenamefont {Santic}}]{Kaufmann1999}%
  \BibitemOpen
  \bibfield  {author} {\bibinfo {author} {\bibfnamefont {U.}~\bibnamefont
  {Kaufmann}}, \bibinfo {author} {\bibfnamefont {M.}~\bibnamefont {Kunzer}},
  \bibinfo {author} {\bibfnamefont {H.}~\bibnamefont {Obloh}}, \bibinfo
  {author} {\bibfnamefont {M.}~\bibnamefont {Maier}}, \bibinfo {author}
  {\bibfnamefont {C.}~\bibnamefont {Manz}}, \bibinfo {author} {\bibfnamefont
  {A.}~\bibnamefont {Ramakrishnan}}, \ and\ \bibinfo {author} {\bibfnamefont
  {B.}~\bibnamefont {Santic}},\ }\href {\doibase 10.1103/PhysRevB.59.5561}
  {\bibfield  {journal} {\bibinfo  {journal} {Phys. Rev. B}\ }\textbf {\bibinfo
  {volume} {59}},\ \bibinfo {pages} {5561} (\bibinfo {year}
  {1999})}\BibitemShut {NoStop}%
\bibitem [{\citenamefont {Koide}\ \emph {et~al.}(2002)\citenamefont {Koide},
  \citenamefont {Walker}, \citenamefont {White}, \citenamefont {Brillson},
  \citenamefont {Murakami}, \citenamefont {Kamiyama}, \citenamefont {Amano},\
  and\ \citenamefont {Akasaki}}]{Koide2002}%
  \BibitemOpen
  \bibfield  {author} {\bibinfo {author} {\bibfnamefont {Y.}~\bibnamefont
  {Koide}}, \bibinfo {author} {\bibfnamefont {D.~E.}\ \bibnamefont {Walker}},
  \bibinfo {author} {\bibfnamefont {B.~D.}\ \bibnamefont {White}}, \bibinfo
  {author} {\bibfnamefont {L.~J.}\ \bibnamefont {Brillson}}, \bibinfo {author}
  {\bibfnamefont {M.}~\bibnamefont {Murakami}}, \bibinfo {author}
  {\bibfnamefont {S.}~\bibnamefont {Kamiyama}}, \bibinfo {author}
  {\bibfnamefont {H.}~\bibnamefont {Amano}}, \ and\ \bibinfo {author}
  {\bibfnamefont {I.}~\bibnamefont {Akasaki}},\ }\href {\doibase
  10.1063/1.1505988} {\bibfield  {journal} {\bibinfo  {journal} {J. Appl.
  Phys.}\ }\textbf {\bibinfo {volume} {92}},\ \bibinfo {pages} {3657} (\bibinfo
  {year} {2002})}\BibitemShut {NoStop}%
\bibitem [{\citenamefont {Lyons}, \citenamefont {Janotti},\ and\ \citenamefont
  {Walle}(2012)}]{Lyons2012}%
  \BibitemOpen
  \bibfield  {author} {\bibinfo {author} {\bibfnamefont {J.~L.}\ \bibnamefont
  {Lyons}}, \bibinfo {author} {\bibfnamefont {A.}~\bibnamefont {Janotti}}, \
  and\ \bibinfo {author} {\bibfnamefont {C.~G. V.~D.}\ \bibnamefont {Walle}},\
  }\href {\doibase 10.1103/PhysRevLett.108.156403} {\bibfield  {journal}
  {\bibinfo  {journal} {Phys. Rev. Lett.}\ }\textbf {\bibinfo {volume} {108}},\
  \bibinfo {pages} {156403} (\bibinfo {year} {2012})}\BibitemShut {NoStop}%
\bibitem [{\citenamefont {Sui}\ and\ \citenamefont {Yu}(2011)}]{Sui2011}%
  \BibitemOpen
  \bibfield  {author} {\bibinfo {author} {\bibfnamefont {Y.-P.}\ \bibnamefont
  {Sui}}\ and\ \bibinfo {author} {\bibfnamefont {G.-H.}\ \bibnamefont {Yu}},\
  }\href {\doibase 10.1088/0256-307X/28/6/067807} {\bibfield  {journal}
  {\bibinfo  {journal} {Chinese Phys. Lett.}\ }\textbf {\bibinfo {volume}
  {28}},\ \bibinfo {pages} {067807} (\bibinfo {year} {2011})}\BibitemShut
  {NoStop}%
\bibitem [{\citenamefont {Reshchikov}\ and\ \citenamefont
  {Morko̧c}(2005)}]{Reshchikov2005d}%
  \BibitemOpen
  \bibfield  {author} {\bibinfo {author} {\bibfnamefont {M.~A.}\ \bibnamefont
  {Reshchikov}}\ and\ \bibinfo {author} {\bibfnamefont {H.}~\bibnamefont
  {Morko̧c}},\ }\href {\doibase 10.1063/1.1868059} {\bibfield  {journal}
  {\bibinfo  {journal} {J. Appl. Phys.}\ }\textbf {\bibinfo {volume} {97}},\
  \bibinfo {pages} {061301} (\bibinfo {year} {2005})}\BibitemShut {NoStop}%
\bibitem [{\citenamefont {Williamson}\ \emph {et~al.}(2004)\citenamefont
  {Williamson}, \citenamefont {Dı́az}, \citenamefont {Bohn},\ and\
  \citenamefont {Molnar}}]{Williamson2004a}%
  \BibitemOpen
  \bibfield  {author} {\bibinfo {author} {\bibfnamefont {T.~L.}\ \bibnamefont
  {Williamson}}, \bibinfo {author} {\bibfnamefont {D.~J.}\ \bibnamefont
  {Dı́az}}, \bibinfo {author} {\bibfnamefont {P.~W.}\ \bibnamefont {Bohn}}, \
  and\ \bibinfo {author} {\bibfnamefont {R.~J.}\ \bibnamefont {Molnar}},\
  }\href {\doibase 10.1116/1.1695335} {\bibfield  {journal} {\bibinfo
  {journal} {J. Vac. Sci. Technol. B}\ }\textbf {\bibinfo {volume} {22}},\
  \bibinfo {pages} {925} (\bibinfo {year} {2004})}\BibitemShut {NoStop}%
\bibitem [{\citenamefont {Moram}\ and\ \citenamefont
  {Vickers}(2009)}]{Moram2009}%
  \BibitemOpen
  \bibfield  {author} {\bibinfo {author} {\bibfnamefont {M.~A.}\ \bibnamefont
  {Moram}}\ and\ \bibinfo {author} {\bibfnamefont {M.~E.}\ \bibnamefont
  {Vickers}},\ }\href {\doibase 10.1088/0034-4885/72/3/036502} {\bibfield
  {journal} {\bibinfo  {journal} {Reports Prog. Phys.}\ }\textbf {\bibinfo
  {volume} {72}},\ \bibinfo {pages} {036502} (\bibinfo {year}
  {2009})}\BibitemShut {NoStop}%
\bibitem [{\citenamefont {Poppitz}\ \emph {et~al.}(2014)\citenamefont
  {Poppitz}, \citenamefont {Lotnyk}, \citenamefont {Gerlach},\ and\
  \citenamefont {Rauschenbach}}]{Poppitz2014b}%
  \BibitemOpen
  \bibfield  {author} {\bibinfo {author} {\bibfnamefont {D.}~\bibnamefont
  {Poppitz}}, \bibinfo {author} {\bibfnamefont {A.}~\bibnamefont {Lotnyk}},
  \bibinfo {author} {\bibfnamefont {J.~W.}\ \bibnamefont {Gerlach}}, \ and\
  \bibinfo {author} {\bibfnamefont {B.}~\bibnamefont {Rauschenbach}},\ }\href
  {\doibase 10.1016/j.actamat.2013.11.041} {\bibfield  {journal} {\bibinfo
  {journal} {Acta Mater.}\ }\textbf {\bibinfo {volume} {65}},\ \bibinfo {pages}
  {98} (\bibinfo {year} {2014})}\BibitemShut {NoStop}%
\bibitem [{\citenamefont {Ryu}\ \emph {et~al.}(2015)\citenamefont {Ryu},
  \citenamefont {Ram}, \citenamefont {Lee}, \citenamefont {Cho}, \citenamefont
  {Lee}, \citenamefont {Kang}, \citenamefont {Kwon}, \citenamefont {Yang},
  \citenamefont {Shin},\ and\ \citenamefont {Woo}}]{yongdeuk2015}%
  \BibitemOpen
  \bibfield  {author} {\bibinfo {author} {\bibfnamefont {S.~R.}\ \bibnamefont
  {Ryu}}, \bibinfo {author} {\bibfnamefont {S.~D.~G.}\ \bibnamefont {Ram}},
  \bibinfo {author} {\bibfnamefont {S.~J.}\ \bibnamefont {Lee}}, \bibinfo
  {author} {\bibfnamefont {H.-d.}\ \bibnamefont {Cho}}, \bibinfo {author}
  {\bibfnamefont {S.}~\bibnamefont {Lee}}, \bibinfo {author} {\bibfnamefont
  {T.~W.}\ \bibnamefont {Kang}}, \bibinfo {author} {\bibfnamefont
  {S.}~\bibnamefont {Kwon}}, \bibinfo {author} {\bibfnamefont {W.}~\bibnamefont
  {Yang}}, \bibinfo {author} {\bibfnamefont {S.}~\bibnamefont {Shin}}, \ and\
  \bibinfo {author} {\bibfnamefont {Y.}~\bibnamefont {Woo}},\ }\href {\doibase
  10.1016/j.apsusc.2015.04.076} {\bibfield  {journal} {\bibinfo  {journal}
  {Appl. Surf. Sci.}\ }\textbf {\bibinfo {volume} {347}},\ \bibinfo {pages}
  {793} (\bibinfo {year} {2015})}\BibitemShut {NoStop}%
\bibitem [{\citenamefont {Sun}\ \emph {et~al.}(2006)\citenamefont {Sun},
  \citenamefont {Selloni}, \citenamefont {Myers},\ and\ \citenamefont
  {Doolittle}}]{Sun2006a}%
  \BibitemOpen
  \bibfield  {author} {\bibinfo {author} {\bibfnamefont {Q.}~\bibnamefont
  {Sun}}, \bibinfo {author} {\bibfnamefont {A.}~\bibnamefont {Selloni}},
  \bibinfo {author} {\bibfnamefont {T.~H.}\ \bibnamefont {Myers}}, \ and\
  \bibinfo {author} {\bibfnamefont {W.~A.}\ \bibnamefont {Doolittle}},\ }\href
  {\doibase 10.1103/PhysRevB.73.155337} {\bibfield  {journal} {\bibinfo
  {journal} {Phys. Rev. B}\ }\textbf {\bibinfo {volume} {73}},\ \bibinfo
  {pages} {155337} (\bibinfo {year} {2006})}\BibitemShut {NoStop}%
\bibitem [{\citenamefont {Romano}\ \emph {et~al.}(2000)\citenamefont {Romano},
  \citenamefont {Northrup}, \citenamefont {Ptak},\ and\ \citenamefont
  {Myers}}]{Romano2000}%
  \BibitemOpen
  \bibfield  {author} {\bibinfo {author} {\bibfnamefont {L.}~\bibnamefont
  {Romano}}, \bibinfo {author} {\bibfnamefont {J.}~\bibnamefont {Northrup}},
  \bibinfo {author} {\bibfnamefont {A.}~\bibnamefont {Ptak}}, \ and\ \bibinfo
  {author} {\bibfnamefont {T.}~\bibnamefont {Myers}},\ }\href {\doibase
  10.1063/1.1318731} {\bibfield  {journal} {\bibinfo  {journal} {Appl. Phys.
  Lett.}\ }\textbf {\bibinfo {volume} {77}},\ \bibinfo {pages} {2479} (\bibinfo
  {year} {2000})}\BibitemShut {NoStop}%
\end{thebibliography}%
\end{document}